\documentclass[a4paper,11pt]{article}
\pdfoutput=1 % if your are submitting a pdflatex (i.e. if you have
\usepackage{jheppub} % for details on the use of the package, please
\usepackage[T1]{fontenc} % if needed
\usepackage{amsmath}
\usepackage{graphicx}
\usepackage{booktabs}
\usepackage{multirow}
\usepackage{xspace}
\usepackage{commands}
\usepackage{subfigure,graphicx}
\usepackage{mathtools}
\usepackage{siunitx}

\usepackage[colorinlistoftodos]{todonotes}

\title{\boldmath Jet Flavour Tagging for Future Colliders with Fast Simulation}

\author[a]{Franco~Bedeschi}
\author[b]{,~Loukas~Gouskos}
\author[b]{and Michele~Selvaggi}

\affiliation[a]{INFN Sezione di Pisa, Italy}
\affiliation[b]{CERN, CH-1211 Geneva 23, Switzerland}

\emailAdd{bed@fnal.gov}
\emailAdd{loukas.gouskos@cern.ch}
\emailAdd{michele.selvaggi@cern.ch}

\abstract{Jet flavour identification algorithms are of paramount importance to maximise the physics potential of future collider experiments. This work describes a novel set of tools allowing for a realistic simulation and reconstruction of particle level observables that are necessary ingredients to jet flavour identification. An algorithm for reconstructing the track parameters and covariance matrix of charged particles for an arbitrary tracking sub-detector geometries has been developed. Additional modules allowing for particle identification using time-of-flight and ionizing energy loss information have been implemented. A jet flavour identification algorithm based on a graph neural network architecture and exploiting all available particle level information has been developed. The impact of different detector design assumptions on the flavour tagging performance is assessed using the FCC-ee IDEA detector prototype.}

\begin{document}
\maketitle
\flushbottom

%%%%%%%%%%%%%%%%%%%%%%%%%%%%%%%%%%%%%%%%%%%%%%%%%%%%%
\clearpage

\section{Introduction}
\label{sec:intro}

Precision measurements of standard model (SM) parameters are key objectives of the physics program of future lepton and hadron  machines~\cite{Abada:2019zxq, CEPCStudyGroup:2018ghi,Baer:2013cma,LC-REP-2013-021,CLICdp:2018cto, Benedikt:2651300}. In particular, the measurement of the Higgs couplings to bottom (\emph{b}) and charm (\emph{c}) quarks, and gluons (\emph{g})~\cite{Asner:2013psa,Thomson:2015jda,Abramowicz:2016zbo, deBlas:2019rxi, An:2018dwb, L.Borgonovi:2642471, Koratzinos:2013ncw}, the Higgs self-coupling~\cite{Mangano:2020sao}  and the precise characterisation of top quark properties, such as the top quark mass~\cite{Seidel:2013sqa} and its electroweak couplings~\cite{Janot:2015yza, Mangano:2015aow} require an efficient reconstruction and identification of hadronic final states. Being able to efficiently identify the flavour of the parton that initiated the formation of a jet, known as jet flavour tagging, is therefore critical for the success of the physics program of future electroweak factories~\cite{Azzi:2021gwg}. The large statistics of hadronic Z boson decays at both lepton circular machines and future hadron machines would provide copious control samples to calibrate jet tagging algorithms in data.

%To this end, powerful identification of the flavour of the parton that initiated the formation of a jet, known as flavour tagging, is critical for the success of the program. Therefore, critical detector components are the pixel and tracking systems which should provide outstanding vertex reconstruction, together with supreme momentum and angular resolution. To meet these requirements, low material detector systems are necessary. Together with the developments on the detector front, sophisticated algorithms are essential to exploit the detector's potential and reach the necessary precision.

Jets originating from \emph{b}  and  \emph{c}  quark decays contain a  \emph{b}  or  \emph{c}  hadron that typically travels a macroscopic distance before decaying into lighter hadrons. Compared to  \emph{b},  \emph{c},  up or down jets (collectively referred to as  \emph{ud}  or light jets in what follows), strange quark (\emph{s}) jets contain a larger fraction of  \emph{s} mesons and baryons. Gluons (\emph{g})  carry a larger colour charge than quarks and thus tend to produce jets with a large particle multiplicity. Quarks have a harder fragmentation function compared to  \emph{g}, which results in a larger fraction of the jet momentum carried by a smaller fraction of the constituents. Jet flavour tagging algorithms aim at identifying these characteristic end-products of the fragmentation and hadronization of the initial parton. 

The first  \emph{b}  and  \emph{c}  quark tagging algorithms were developed at LEP~\cite{Abdallah:2002xm,Proriol:1950599} and the Tevatron~\cite{D0:2010zho,Freeman:2012uf}. These algorithms typically rely on the detector capability to identify and measure charged tracks with a significant displacement ($\ctau \sim 500~(150)~\um$) from the beam axis originated from long-lived B (D) meson weak decays. On the other hand, tracks from the charged hadrons produced in  \emph{ud}  quark decays feature a small distance at closest approach to the interaction point. Therefore, in the case of  \emph{b}  and  \emph{c}  tagging, tracks are typically clustered to reconstruct possible secondary vertices (SVs). However,  \emph{c}  tagging is more challenging than  \emph{b}  tagging, due to its properties laying between the  \emph{b}  and \emph{ud} or  \emph{g}  jets.  
The track multiplicity and the mass of the SV (expected to be large for heavy flavour jets), together with the presence of a non-isolated electron or muon indicating a semi-leptonic heavy flavour decay, are also used as discriminating variables in traditional heavy quark tagging algorithms. Such taggers, widely used in the early days of current LHC experiments~\cite{ATLAS-CONF-2010-042,ATLAS-CONF-2010-070,ATLAS-CONF-2010-091,CMS-PAS-BTV-11-004} are implemented by directly applying a selection on a combination of the tracks and SV properties, by constructing a likelihood ratio or a multi-variate discriminant based on a set of jet--level properties. 

Recently, a new generation of advanced machine learning based jet tagging algorithms has been developed~\cite{ATLASBTag2016, CMS:JME, ATL-PHYS-PUB-2017-003, Bols:2020bkb}, bringing more than an order of magnitude improvement in background rejection compared to the traditional approaches in heavy flavour and  \emph{g}  tagging.  
Three are the primary reasons for this success. 
First, significant advancements in the architecture of the neural networks used, as well as new jet representations that allow to better capture the jet properties have been achieved. 
Second, these algorithms exploit directly low-level information, e.g., from reconstructed particles (as in the Particle-Flow algorithm~\cite{CMS:2017yfk}) or even reconstructed hits, compared to traditional methods. This allows to explore in much more depth the true potential of the detectors and the event reconstruction, and also better capture the jet properties compared to algorithms relying on jet-level observables. Moreover, the nature of each of the jet constituents, via particle identification techniques (PID), is expected to provide an additional useful handle in discriminating between different jet species. Powerful particle identification capabilities based on ionisation energy loss (via dE/dx or cluster counting), or via precise time-of-flight measurements, are expected to be highly beneficial for jet flavour tagging, in particular for  \emph{s}  tagging where the identification of charged kaons is crucial. 
Finally, the developments in computing, e.g., graphics processing units, and the availability of very large Monte Carlo simulated and collision data samples, were critical for the development of these advanced methods. 

In this paper we present a general framework for building a jet flavour tagging at future colliders using fast detector simulation and state-of-the art machine learning techniques. A major goal of the present work has been to allow for the evaluation of the impact of specific detector design options on the jet flavour tagging performance (and in turn on the physics potential) in an efficient yet precise way. To this end, we have implemented two key additions to the official \delphes\ fast simulation framework. The \trackcovMod~\cite{tc_module}, described in Section~\ref{sec:ftc}, which allows for a simple definition of a tracker geometry, and the fast simulation and the reconstruction of the parameters and covariance matrix of charged particles tracks. The \tofMod~\cite{tof_module} and \ccMod~\cite{cc_module}, described in Section~\ref{sec:pid}, open up the possibility to model particle identification in \delphes. Section~\ref{sec:flavour} describes the input observables and the implementation of the jet flavour identification algorithm. The tagging algorithm is based on \pnet~\cite{Qu:2019gqs}, using state-of-the-art jet representation and a graph neural network (GNN) architecture. The performance of the algorithm is evaluated using one of the FCC-ee/CEPC baseline detector concepts, the IDEA~\cite{Bedeschi:2021nln, Abada:2019zxq,delphes_card_idea} detector. Variations around the baseline using Higgs decays taken from a Higgsstrahlung sample at $\sqrt{s} = 240$~GeV are discussed. Finally, a discussion of the results, together with limitations of the current approach and perspectives for future work are presented in Section~\Ref{sec:disc}.

%\begin{figure}[ht!]
%  \centering
%    \includegraphics[width = 0.95\linewidth]{Fig/diagrams.pdf}
%    \caption{a caption}
%  \label{fig:diagrams}
%\end{figure}

%%%%%%%%%%%%%%%%%%%%%%%%%%%%%%%%%%%%%%%%%%%%%%%%%%%%%

\section{Fast Tracking simulation}
\label{sec:ftc}
%%%%%%%%%%%%%%%%%%%%%%%%%%%%%%%%%%%%%%%%%%%%%%%%%%%%

%The track impact parameter resolution is the most important factor in the performance heavy quark flavour identification and can be achieved with high granularity, minimal material budget and proximity of the first pixel layer to the beam-pipe. In  we describe the implementation of that fast tracking simulation 

%the response of different tracking design options, whereas Section~\ref{sec:pid} presents the two particle identification techniques used to access the performance of the vertex and tracking systems. 

The tracking system is a major part of modern detectors for high energy physics experiments and arguably the most relevant for jet flavour tagging since it is responsible for reconstructing and identifying charged particles. The design of this system, its optimization, and the evaluation of its performance on many specific physics benchmarks is a fundamental step in the planning of future experiments. To this end, we have developed and included in \delphes, a versatile and modular framework to easily study different detector configurations, and provide for each of those a fast simulation of the tracking performance. The corresponding module is named \trackcovMod~\cite{tc_module}. In this section we present the general implementation of the algorithm, while technical details on the speed optimisation and randomisation can be found in Appendix~\ref{app:trkopt}.

%The main ingredients for such a framework are:

%\begin{itemize}
%\item[i.] {a simple definition of the geometry;}
%\item[ii.]{the calculation of the ``true'' track parameters;}
%\item[iii.] {the calculation of the covariance matrix of those parameters from the assumed detector geometry, resolution and the materials present inside the tracking volume;}
%\item[iv.]{randomization of the ``reconstructed'' track parameters according to the covariance matrix;}
%\item[v.]{speed up any time consuming part of the previous processes;}
%\item[vi.]{implementation of fast tracking in the context of a program that performs simulation of physics processes as provided by many existing particle generators.}
%\end{itemize}

While various attempts to calculate the track resolution analytically have been made (see for instance~\cite{Drasal:2018zij}), they usually make highly simplifying assumptions such as equal spacing and equal detector resolution, that make them unsuitable to use for a realistic combined tracking system. The tracking system geometry is described in terms of  \emph{layers}. Only two types of layer geometries are considered: cylinders coaxial with the beam axis and planar disks orthogonal to the beam axis ($z$-direction). Each layer can be either associated to a measurement with a given resolution or else be just included to describe passive material in the system.  An accurate description of the material inside the tracking volume is important to estimate appropriately the contribution of multiple coulomb scattering to the track resolution. Several measurement geometries are allowed: axial or stereo strips and wires, and pixels.

%In the following an alternative approach that combines simplicity and flexibility is described.

%\subsection{Geometry description}
%\label{sec:ftc_geom}

%The full geometrical description is included in the framework~\cite{ref2code} in a way that enables the study of different detector configurations or even concepts in a simple manner. (I think this paragraph needs re-writing - reads very colloquial)

%\subsection{Track covariance matrix generation}
%\label{sec:ftc_cov}

The tracking system is located inside the solenoid magnet generating a constant field, $B$, directed parallel to the $z$-direction. With these assumptions, charged tracks follow a helix trajectory that is described with a set of five parameters: $\vec{\alpha}=(D,\, \varphi_0,\,C,\,z_0,\,\lambda)$. These parameters are defined in the point of closest approach (PCA) of the track to the $z$-axis; $D$ is the signed transverse distance of the PCA from the $z$-axis, $\varphi_0$ is the track azimutal angle, $C$ is the signed half curvature, $z_0$ the $z$-coordinate of the PCA and $\lambda$ the cotangent of the track polar angle. Given a charged particle originating at $\vec{x}$ with momentum $\vec{p}$ and charge $Q$, the parameters $\vec{\alpha}$ are uniquely defined, as is the associated trajectory.
	
During its motion the charged particle will cross some of the layers described in the geometry. At each crossing the particle will undergo small random changes of its direction due to multiple scattering and, in the case of measurement layers, a generalized coordinate, $d^*$, will be measured with an uncertainty given by the specific detector resolution. The track parameters are reconstructed from the measured coordinates by minimizing the following $\chi^2$ with respect to the track parameters $\vec{\alpha}$. The $\chi^2$ is defined as:
\begin{equation}
\label{eq:chi2}
\chi^2 = ~(\vec{d}-\vec{d}^*)^t\,S^{-1}(\vec{d}-\vec{d}^*),
\end{equation}
where $\vec{d}^*$ is the array of measured coordinates and $\vec{d}$ that of predicted coordinates, that can be computed from the track parameters $\vec{\alpha}$ and the geometry of each measurement layer. $S$ is the covariance matrix of all the measurements and includes contributions from the detector resolution and from the multiple scattering. The superscript $t$ indicates the transpose of a vector or a matrix.  Assuming $\vec{d}_0$ to be the array of predicted coordinates for which the $\chi^2$ is minimized, then for small variations of the track parameters relative to this minimum, $\delta\vec{\alpha}$, we have:
\begin{equation}
\label{eq:d_lin}
\vec{d} \simeq  \vec{d}_0+\frac{\partial \vec{d}}{\partial \vec{\alpha}}\delta\vec{\alpha}
= \vec{d}_0+A\delta\vec{\alpha}, 
\end{equation} 
where $A$ is the derivative matrix. Including Equation.~(\ref{eq:d_lin}) in Equation.~(\ref{eq:chi2}) we obtain:
\begin{equation}
\label{eq:chi2lin}
\chi^2 =  (\vec{d}_0-\vec{d}^*+A\delta\vec{\alpha})^t\,S^{-1}(\vec{d}_0-\vec{d}^*+A\delta\vec{\alpha}).
\end{equation}
Differentiating with respect to the track parameters we obtain the track parameter covariance matrix, $C$:
\begin{equation}
\label{eq:covmat}
C^{-1}=\frac{1}{2}\frac{\partial^2 \chi^2}{\partial\vec{\alpha} \partial\vec{\alpha}}
= A^t\,S^{-1}A.
\end{equation}
This equation highlights the key ingredients to estimate the track covariance matrix; the derivative matrix and the covariance matrix of the measurements. The former is straightforward and can be derived for every type of measurement given the track equation. The latter requires the combination of two elements: the intrinsic detector resolution and the multiple scattering contribution, as shown in the following equation:
\begin{equation}
\label{eq:cov_meas}
S_{ij} = \sigma_i^2\,\delta_{ij}+M_{ij}, 
\end{equation}
where the indices $i$ and $j$ identify the measurement layers, $\sigma_i$ is the detector resolution for  layer $i$  and $M_{ij}$ is the multiple scattering contribution. The $M_{ij}$ includes contributions from all scattering layers below the smallest of the two indices, as shown in the following equation:
\begin{equation}
\label{eq:mscat}
M_{ij} = \sum_{1\le k<\mathrm{min}(i,j)} (L_i-L_k)(L_j-L_k) \theta_k^2(i,j),
\end{equation}
where $L_i$ is the distance traveled by the track to the layer $i$ and $\theta_k(i,j)$ the standard deviation of the multiple scattering angle generated by layer $k$ after correcting for projection factors specific for layers $i$ and $j$.

\begin{figure}[htb]
    \begin{center}
        {\includegraphics[scale=0.6]{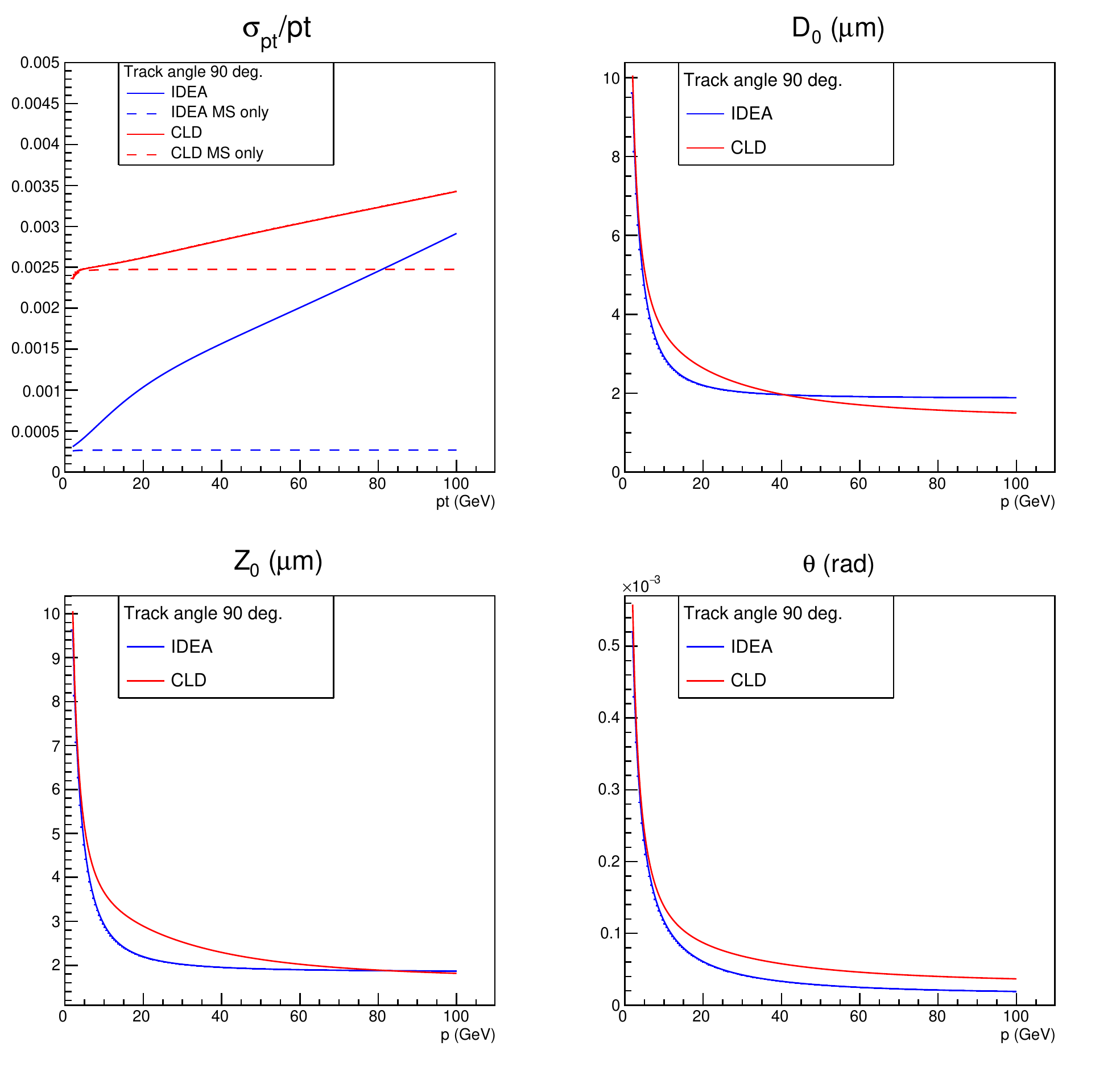}}
        \caption{Track parameter resolution for the IDEA and CLD detector concepts for FCC-ee~\cite{Abada:2019zxq}. The dashed lines in the top left plot show the multiple scattering contribution.}
\label{fig:resol}
    \end{center}
\end{figure}

Once $A$ and $S$ have been determined for a given track, the parameter covariance matrix can be computed analytically using Equation.~(\ref{eq:covmat}).  The obtained resolution as a function of momentum of the track parameters (\pt,\,$D$,\,$z_0$,\,$\theta$)  for two different reference detector configurations proposed for a future $e^+e^-$ collider is shown in Figure~\ref{fig:resol}.

\section{Particle identification}
\label{sec:pid}
%%%%%%%%%%%%%%%%%%%%%%%%%%%%%%%%%%%%%%%%%%%%%%%%%%%%%

Particle identification techniques can play a major role in the identification of the jet flavour. In particular, as will be discussed in Section~\ref{sec:flavour},  \emph{s}  jets contain a significant fraction of charged kaons (\kpm) compared to  \emph{u} or  \emph{d} jets that are mostly composed of charged pions (\pipm). Given that the performance of such algorithms heavily depends on explicit detector design choices, it is crucial to be able to first simulate appropriately the detector response and then to implement such particle identification algorithms. 

Two complementary particle identification techniques have been included in the \delphes\ fast-simulation. The precise measurement of the time of arrival of tracks in the outermost part of the tracking volume, together with the momentum and the path length, provide an indirect measurement of the particle mass via the well known time-of-flight method. This method has been implemented in the \tofMod\ module~\cite{tof_module}. The cluster counting method, $dN/dx$, implemented in the \ccMod\ module~\cite{cc_module}, consists in counting the multiplicity of the primary ionization clusters produced along the track in gaseous detectors, which together with the particle momentum, can also be used to infer the particle mass. In this section we discuss the implementation of these two methods within the simulation framework. 

\clearpage

\subsection{Time-of-flight}
\label{subsec:tof}

The time-of-flight (\tof) of a particle can be expressed as:

\begin{equation}
\label{eq:tof_main}
\tof \equiv \tf - \tv = \frac{L}{\beta} = \frac{L \sqrt{p^2 + m^2}}{p} =  \frac{L E}{\sqrt{E^2 - m^2}}, 
\end{equation}

where \tf\ is the measured time after propagation, \tv\ is the particle time of production at vertex, $L$ is total path length, and $p$, $E$ and $m$ are the momentum, energy and mass of the particle, respectively. Provided that the quantities $L$ and $p$ (or $E$) and \tv\ can be measured, the measurement of \tof\ provides an estimate for the particle mass and thus a powerful handle for particle identification. 

For charged particles the reconstructed mass is given by: 

\begin{equation}
\label{eq:mtof_charged}
\mtof^{(c)} = p \sqrt{(\frac{\tof}{L})^2 - 1}, 
\end{equation}

\begin{figure}[ht]
    \centering
     \subfigure[]{\includegraphics[width = 0.47\linewidth]{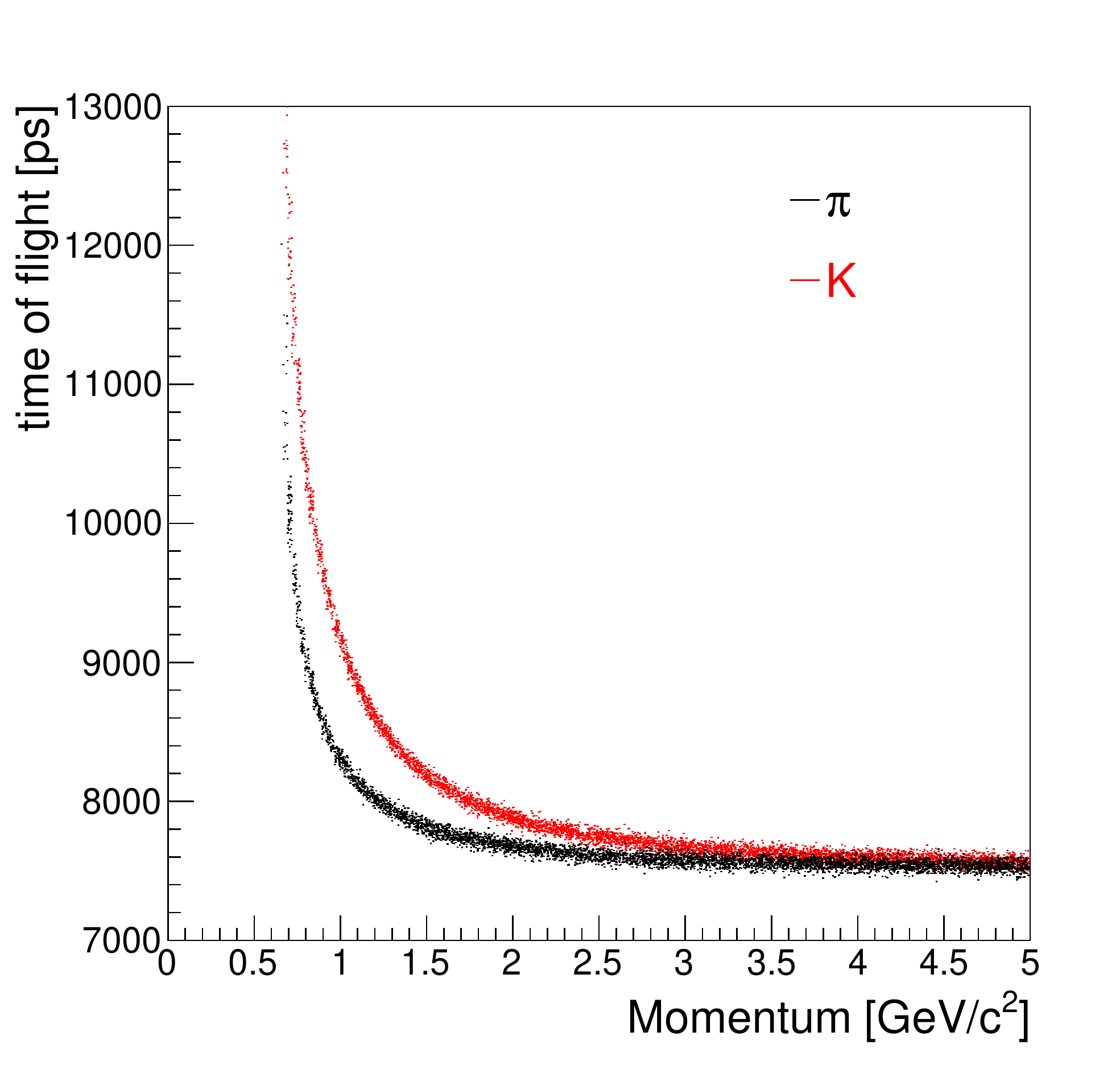}}
     \subfigure[]{\includegraphics[width = 0.47\linewidth]{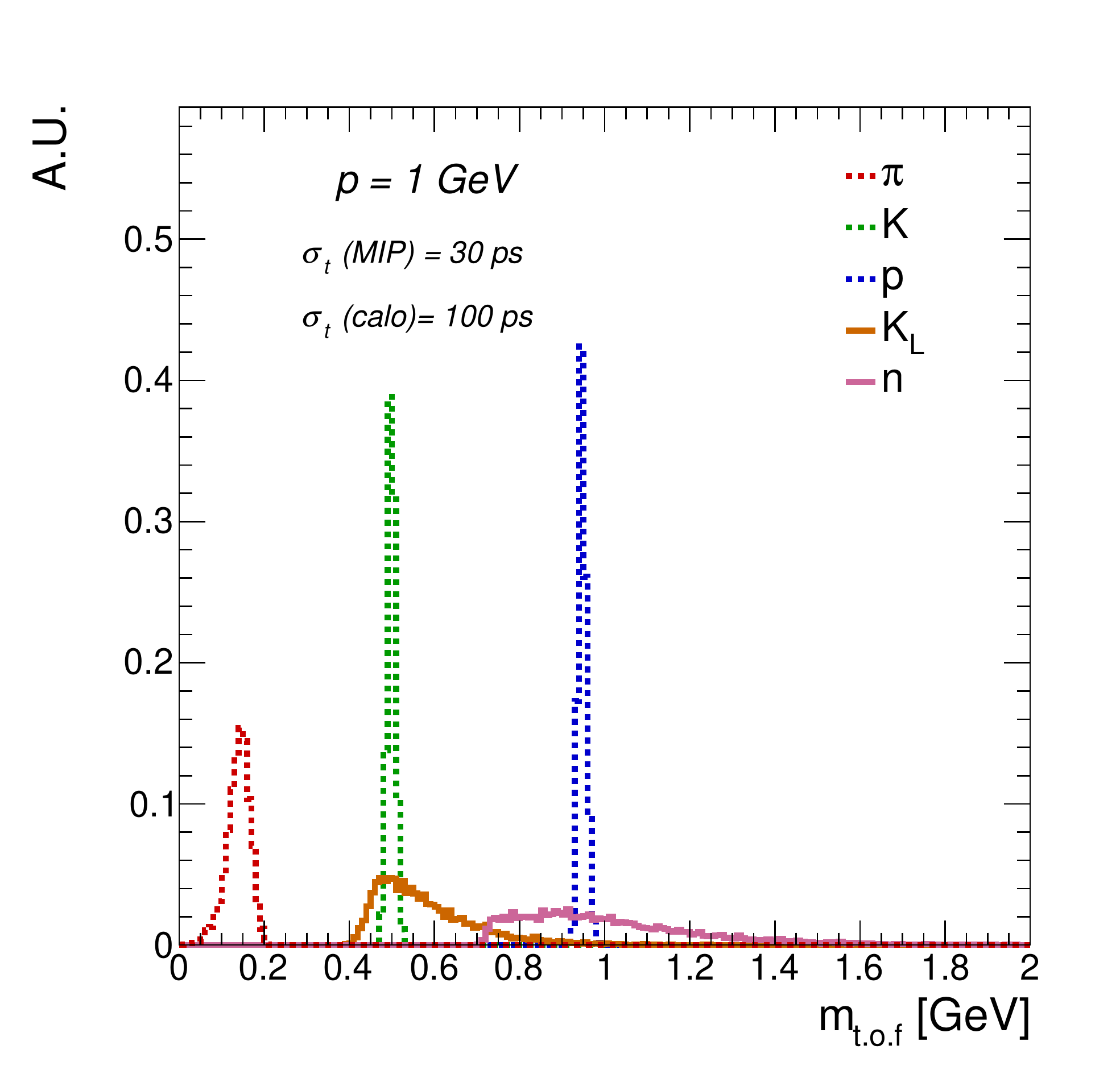}}
    \caption{\label{fig:tof} (a) Time-of-flight for \kpm\ and \pipm\ track at $\theta = 90\,^\circ$ as a function of momentum in the IDEA detector drift chamber. (b) Reconstructed \mtof\ for \kpm, \pipm, \kl, protons and neutrons.}  
\end{figure}

The initial position (and therefore $L$) and the particle momentum $p$ are reconstructed by means of the inner/outer tracking system, and simulated with the procedure described in Section~\ref{sec:ftc}. The time of a particle production at vertex \tv\ can be estimated indirectly, with the following procedure. Assuming that the beamspot has a small time (\sbz) and longitudinal (\sbt) spread compared to the precision of the timing measurement device, the time of the primary vertex can be simply taken as $\tpv=0$. However, if the particle originates from a highly displaced vertex (e.g. from \Ks\ or $\Lambda$.), assuming $\tv=0$ can lead to a severe over-estimate of \tof. A more accurate estimate for the vertex time corresponds to $\tv = \frac{r_\text{V}}{\beta_\text{V}}$, where $r_\text{V}$ is the distance of the vertex to the origin and $\beta_\text{V}$ is the vertex velocity, computed from its outgoing particles. In the current study we assume we are able to reconstruct the initial time at vertex perfectly and therefore we take the initial time from Monte Carlo simulation. The time of flight distribution of charged Kaons and Pions emitted at 90$\,^\circ$ is shown in Figure~\ref{fig:tof} (a), assuming a 30 ps timing resolution, which allows for an efficient $K/\pi$ discrimination for momenta $p<3.5$~GeV.

For neutral particles the mass can be reconstructed from the energy measurement provided by the calorimeters: 

\begin{equation}
\label{eq:mtof_charged}
\mtof^{(n)} = E \sqrt{ 1 - (\frac{L}{\tof})^2}, 
\end{equation}

At low momenta, where the time-of-flight method is expected to provide good identification capabilities, the calorimetric energy measurement is sub-optimal and leads to poor \mtof\ resolution for neutral particles compared to charged particles. Moreover, the vertex time determination is inaccessible for neutral particles. The assumption $\tpv=0$ for all neutral particles leads to an additional uncertainty on the \mtof\ estimate. The reconstructed \mtof\ for \kl\ and neutrons is shown in Figure~\ref{fig:tof} (b).

\subsection{Cluster counting}
\label{subsec:clustercounting}

The cluster counting technique is expected to provide  improved particle identification relative to the more commonly used $dE/dx$ methods in large drift chambers or TPCs~\cite{Walenta:1979ut, Caron:2013gca}. In addition it does not require the tuning of truncated mean algorithms to suppress the large Landau tails present in the $dE/dx$ distribution. The number of ionization clusters per unit length is obtained from a very detailed simulation program, Heed++~\cite{Smirnov:2005yi}, now fully integrated into Garfield++~\cite{Veenhof:1998tt}. An array of number of ionization clusters per unit length for several values of $\beta\gamma$ is obtained from Garfield and used to interpolate the average cluster density. The total mean number of clusters is found by multiplying for the track length in the chamber. Finally the observed cluster number is obtained by extraction over a Poisson distribution with that mean. Four common gas options are available: pure Helium or Argon, He 90\% + Isobutane 10\%, Argon 50\% + Ethane 50\%. This library can be easily extended if needed to a larger collection of gas mixtures.

\begin{figure}[htb]
    \centering
     \subfigure[]{\includegraphics[width = 0.47\linewidth]{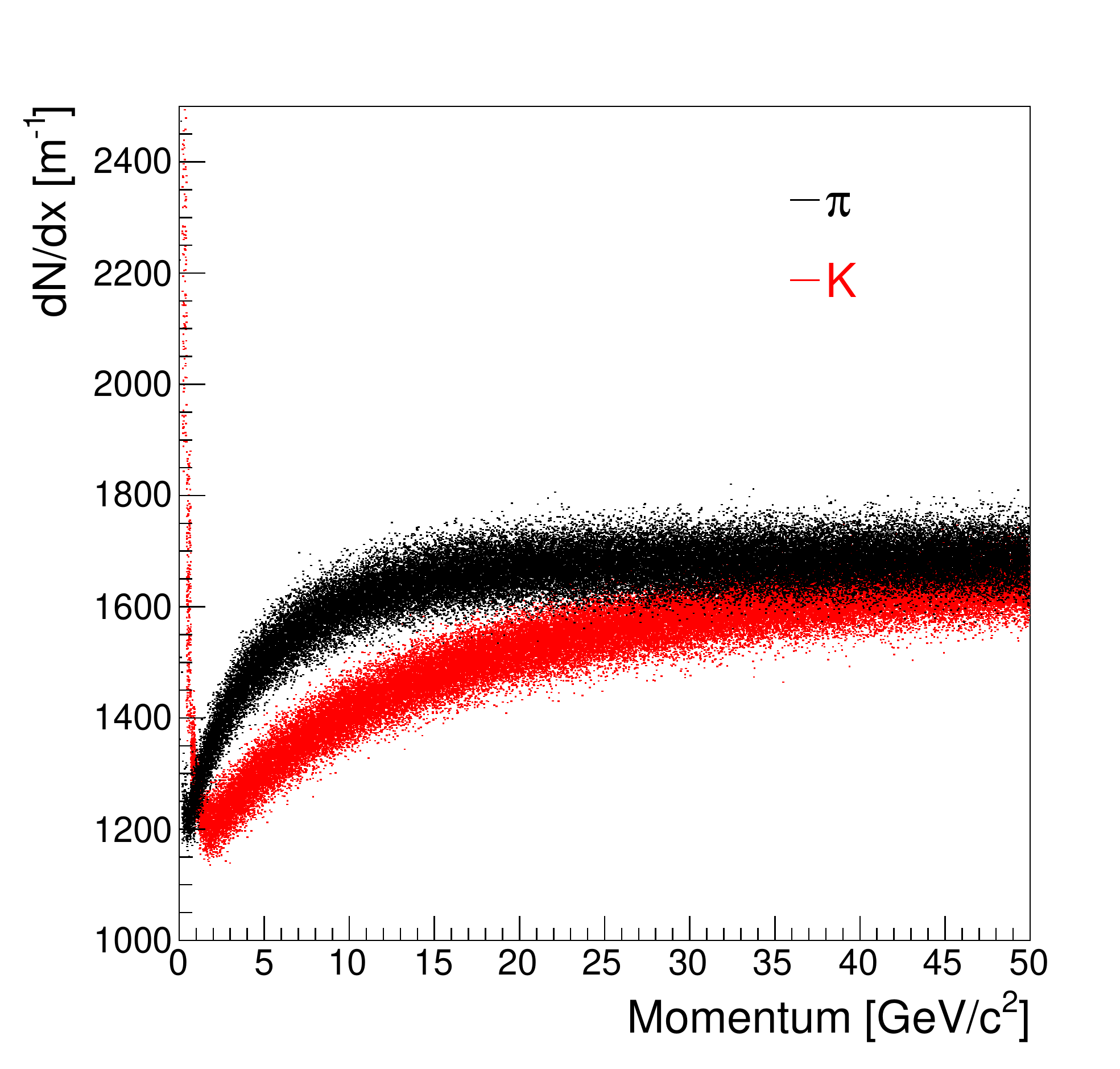}}
     \subfigure[]{\includegraphics[width = 0.47\linewidth]{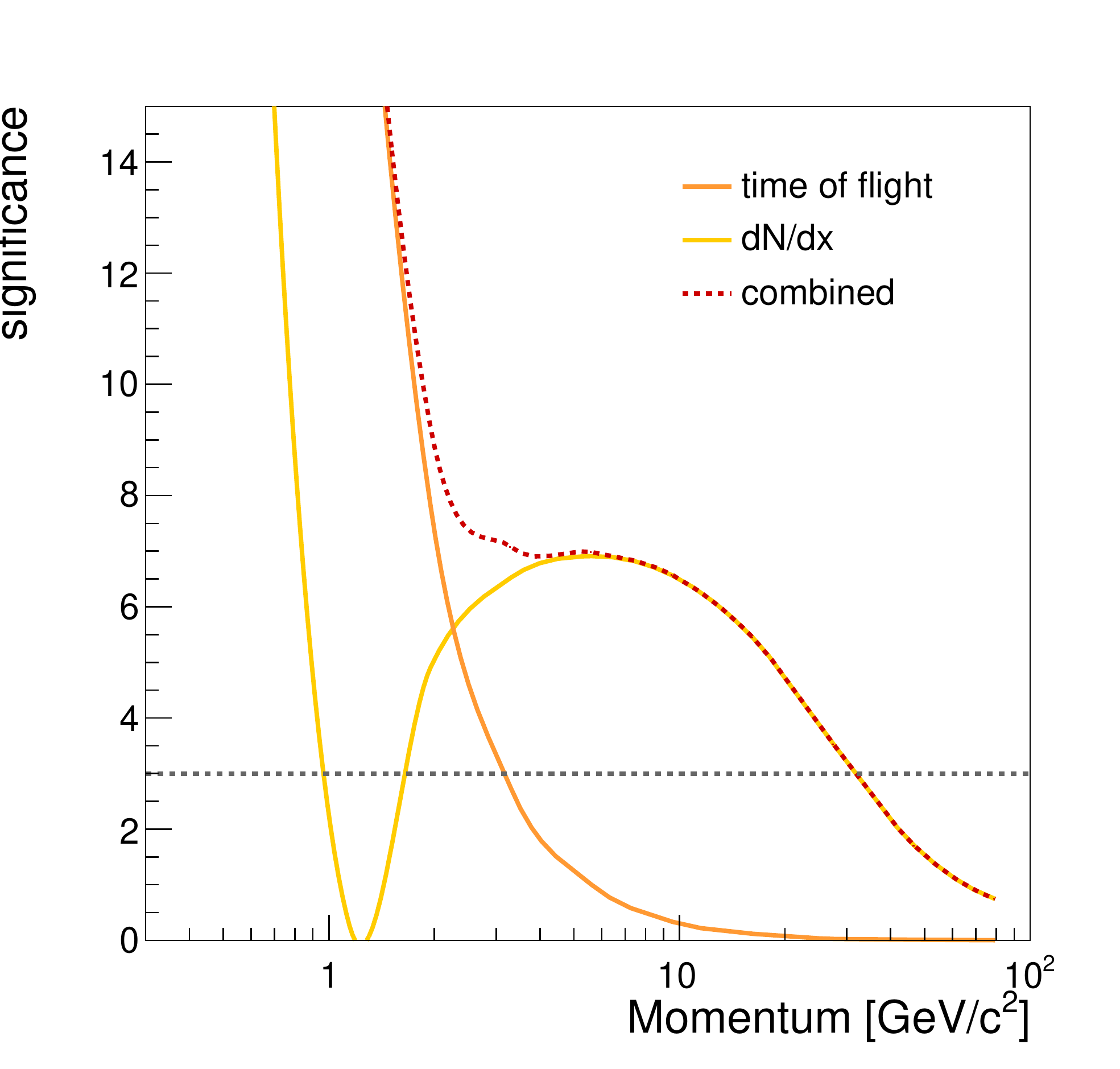}}
    \caption{\label{fig:dndx_comb} (a) Number of cluster distribution of charged pions and kaons for 90$\,^\circ$ tracks in the IDEA detector drift chamber as function of momentum; (b) $K/\pi$ separation in number of $\sigma$ as a function of the particle momentum using the $dN/dx$ and time-of-flight methods.}  
\end{figure}

In Figure~\ref{fig:dndx_comb} (a) the potential for $K/\pi$ separation is shown for a He 90\% + Isobutane 10\% mixture over a wide range of momenta ($2<p<30$~GeV) . The combination of the cluster counting and time-of-flight techniques is displayed in Figure~\ref{fig:dndx_comb} (b) and shows an efficient separation of \kpm\ / \pipm\ separation ($\ge 3\sigma$) for momenta $p<30$~GeV.

\section{Jet flavour identification}
\label{sec:flavour}

In this Section a novel jet tagging algorithm is presented. The jet flavour discrimination uses reconstructed observables at the level of the jet constituents. For simplicity, the jet flavour discriminant is built and evaluated using \ee\ collisions reconstructed with the IDEA detector concept and will thus be referred as \pnetee. While the obtained performance is specific to the clean \ee\ environment and the explicit detector specifications, the inputs and the construction of the discriminant itself are general. We first discuss the event generation and reconstruction details, then introduce the particle-level input observables and the architecture of the neural network discriminant. Finally we address the tagger performance and its robustness with respect to different detector choices.

\subsection{Simulated data}
\label{subsec:input}

The simulated sample consists of \eezh\ events produced at a center of mass energy \sqrtsfcceeh. The Higgs bosons decay to \hgg\ or \hqq, where $q=(u,d), s, c, b$ with relative fraction as expected for a SM Higgs boson with $m=125$ GeV,  whereas the Z bosons always decay to a pair of neutrinos. The hard scattering process is generated with \MGAMC~\cite{Alwall:2014hca}, while \py~\cite{Sjostrand:2014zea} is used for modeling the decay, parton-shower and hadronisation processes. Five different samples, corresponding to each jet flavour category ( \emph{ud},  \emph{s},  \emph{c},  \emph{b},  \emph{g} ) containing $10^6$ events each (or equivalently $2 \times 10^6$ jets) are used for the training.
Final state particles are reconstructed with the \delphes\ PF algorithm. In particular, charged particles are reconstructed using the latest \trackcovMod\ module described in Section~\ref{sec:ftc}, and the time-of-flight and number of ionisation clusters per unit length (\dndx), are reconstructed using the \tofMod\ and \ccMod\ modules, described in Sections~\ref{subsec:tof} and~\ref{subsec:clustercounting}, respectively. Neutral particles (photons and neutral hadrons) are reconstructed by the PF algorithm implemented in the \drMod\ module~\cite{dr_module}. The time-of-flight (and corresponding reconstructed mass \mtof) of neutral hadrons is also included and assumes a 100 ps resolution, as opposed to 30 ps assumed for charged particles. The baseline simulation setup assumes the nominal IDEA detector concept~\cite{Bedeschi:2021nln, delphes_card_idea}. Jets are clustered with the \fastjet~\cite{Cacciari:2011ma} package using the \ee\ generalized \kt\ algorithm~\cite{Cacciari:2008gp, CATANI1991432} with parameter $p=-1$. 

\subsection{Input features}
\label{subsec:inputfeatures}
The jet constituents in the form of PF candidates are used as inputs to the \pnetee\ algorithm. For each PF candidate we define a set of input observables (features) that are summarized in Table~\ref{tab:pnetinputs}. The first set of inputs, denoted as  \emph{kinematics}, uses features derived from the 4-momentum of each jet constituent. These include the energy measurement of the constituent relative to the jet energy and the direction of the jet constituents relative to the jet momentum. %It also includes an observable denominated $\ptd = \sqrt{\sum {p_{\textrm{T}}^{2}(i)}}/\sum {p_{\textrm{T}}(i)}$ \MS{do we really need ptd ? }. The upper row shows an example of jet-level observables which rely only on kinematic  information of the jet particles. 
The second set of features, labelled as  \emph{displacement}, includes observables related to the longitudinal and transverse displacement of the jet constituents which are more relevant to identify jets originating from the hadronization of the \emph{b} and \emph{c} quarks. 
Finally, the third set of inputs, labelled as  \emph{identification}, refers to the nature of each particle using the PF reconstruction and the particle identification (PID) algorithms presented in Section~\ref{sec:pid}. 

\begin{table}[!h]
    \caption{\label{tab:pnetinputs}Set of input variables}
    \centering
    \begin{tabular}{cc} \hline
    Variable & Description \\ \hline \hline
        \multicolumn{2}{c}{Kinematics} \\ \hline
\erel & energy of the jet constituent divided by the jet energy \\
%\sth & sin of the angle between the constituent momentum and the jet momentum \\
%\cth & cos of the angle between the constituent momentum the jet momentum \\
%\ptd\ & fragmentation function (defined in text) \\
\thetarel & polar angle of the constituent with respect to the jet momentum\\
\phirel & azimuthal angle of the constituent with respect to the jet momentum\\

\hline \hline

\multicolumn{2}{c}{Displacement} \\ \hline

$d_{xy}$ & transverse impact parameter of the track \\
$d_z$ & longitudinal impact parameter of the track \\
\sipIIDval &  signed 2D impact parameter of the track \\
\sipIIDsig &  signed 2D impact parameter significance of the track \\
\sipIIIDval &  signed 3D impact parameter of the track \\
\sipIIIDsig &  signed 3D impact parameter significance of the track \\
\distval & jet track distance at their point of closest approach \\ 
\distsig & jet track distance significance at their point of closest approach  \\
\covmatrix & covariance matrix of the track parameters\\  
\hline \hline

\multicolumn{2}{c}{Identification} \\ \hline
$q$ & electric charge of the particle\\ 
\mtof & mass calculated from time-of-flight \\
$dN/dx$ & number of primary ionisation clusters along track \\
\texttt{isMuon} & if the particle is identified as a muon \\
 \texttt{isElectron} & if the particle is identified as an electron \\
 \texttt{isPhoton} & if the particle is identified as a photon \\
 \texttt{isChargedHadron} & if the particle is identified as a charged hadron \\
 \texttt{isNeutralHadron} & if the particle is identified as a neutral hadron
 \\ \hline \hline
 \end{tabular}
\end{table}

%\MS{maybe move ptrel pand ptpar to kinematics?} \LG{Agree - are these input features the final ones?}
%\MS{Lots of redundancies in the list of vars. I am now retraining with the minimal set that is in the table}

\begin{figure}[t]
     \centering 

     \subfigure[]{\includegraphics[width = 0.47\linewidth]{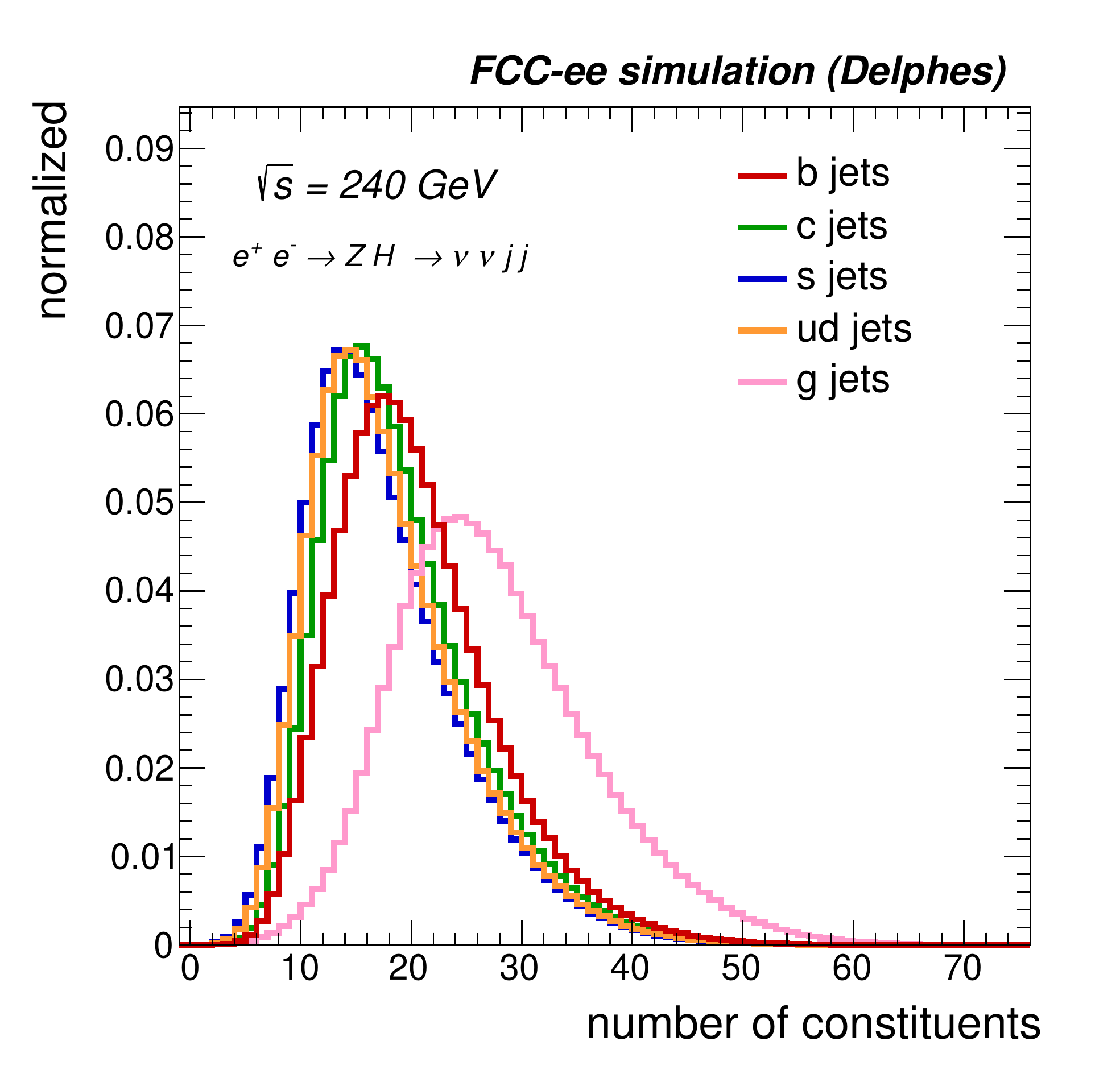}}
     \subfigure[]{\includegraphics[width = 0.47\linewidth]{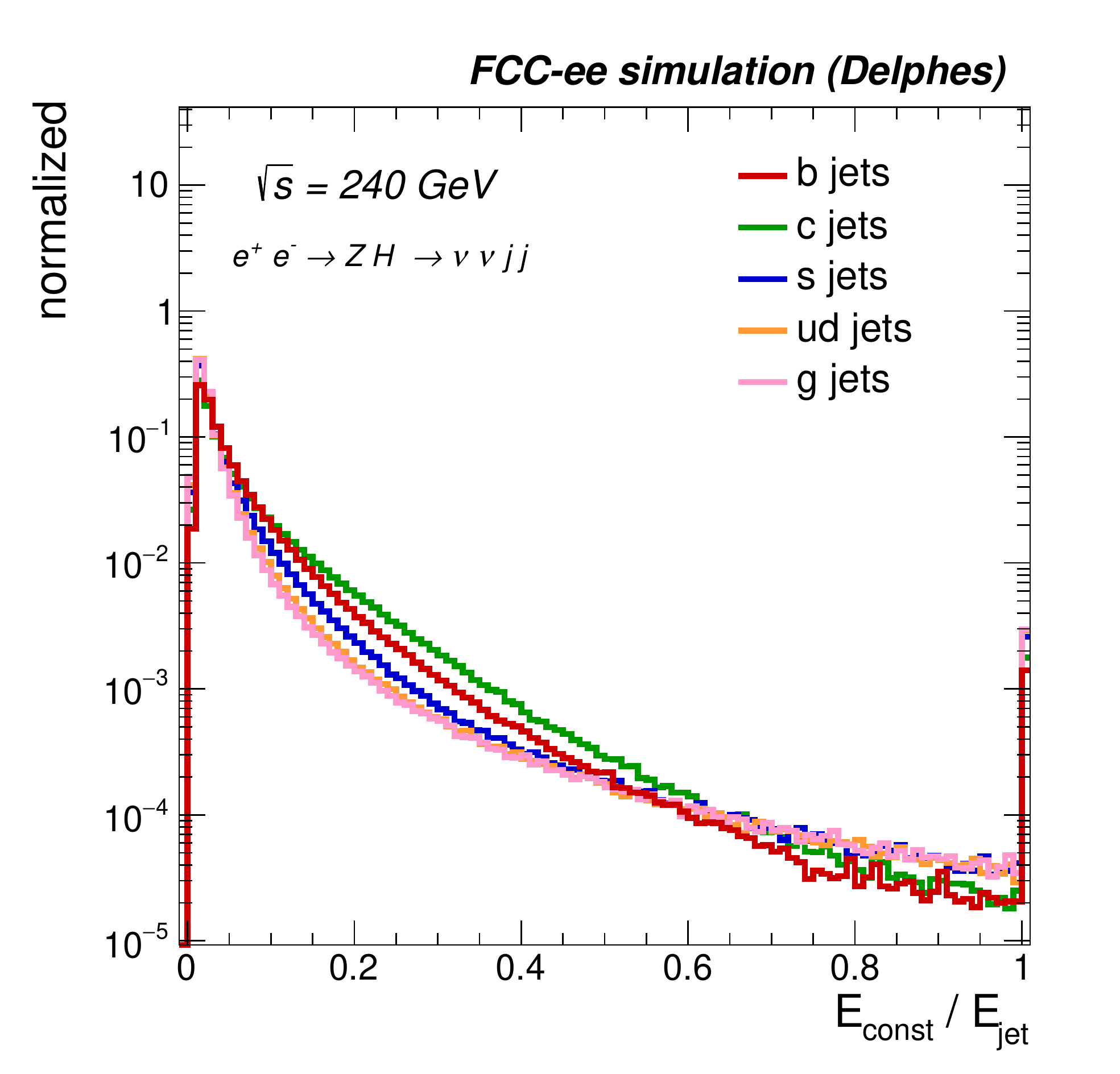}}
     \subfigure[]{\includegraphics[width = 0.47\linewidth]{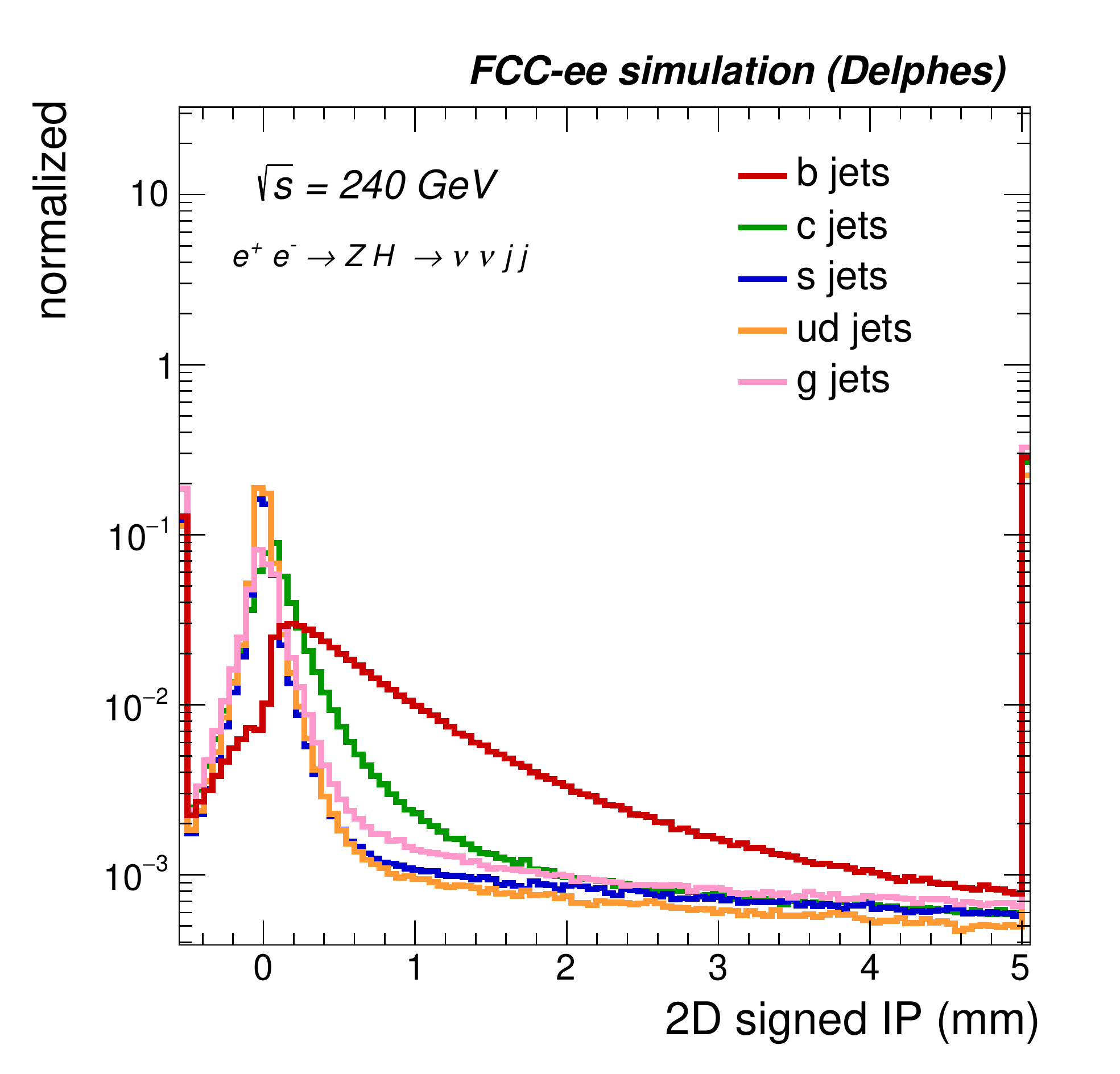}}
     \subfigure[]{\includegraphics[width = 0.47\linewidth]{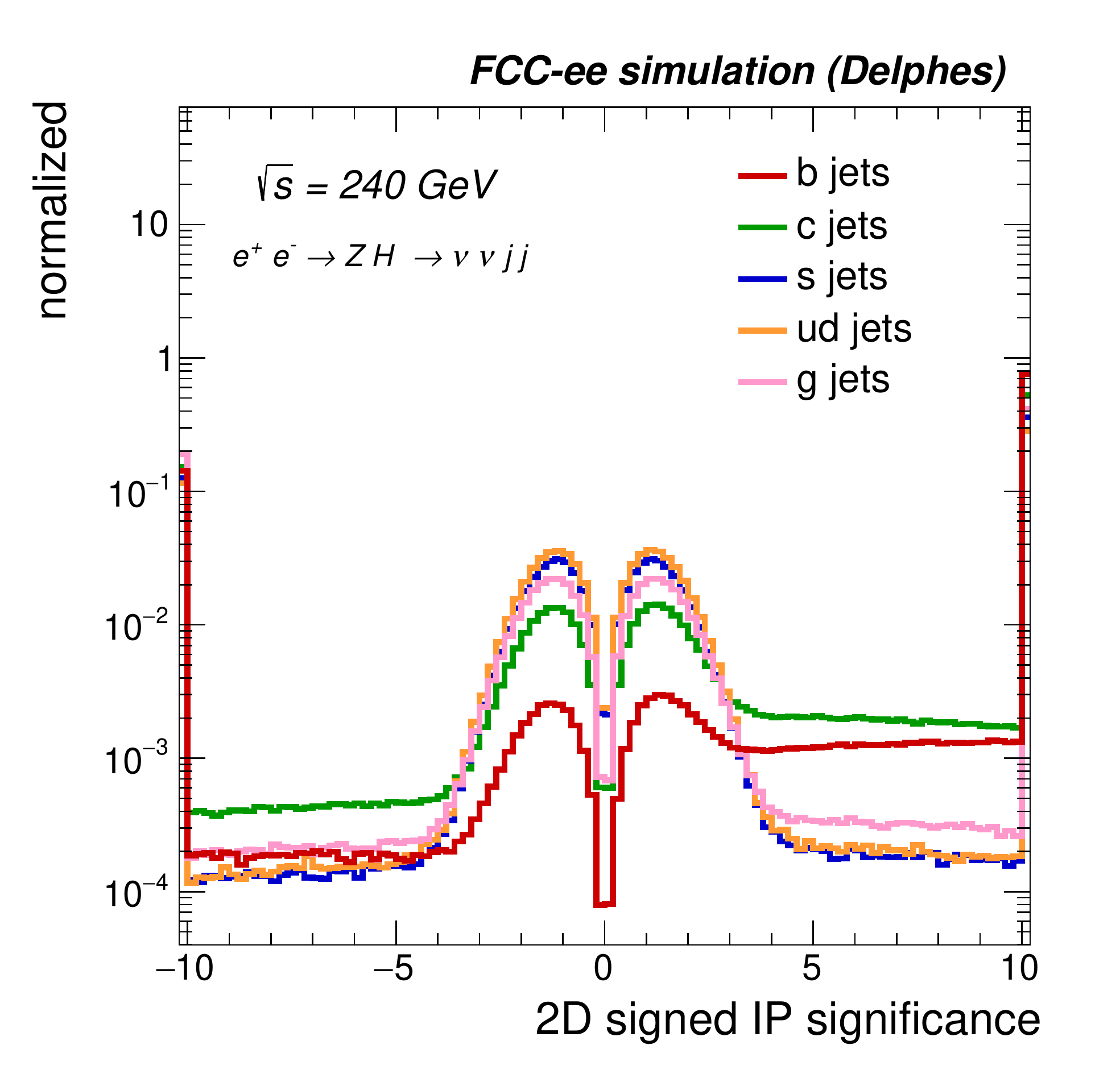}}

        \caption{\label{fig:input_comparison_idea_flavours}Shape comparison of a set of representative observables relevant for jet flavour identification. The different colors correspond to different jet flavours. The FCC-ee IDEA detector concept is used.}
\end{figure}

 The total number of reconstructed jet constituents, shown in Figure~\ref{fig:input_comparison_idea_flavours}(a), is typically larger for  \emph{g}  jets compared to quark jets due to their different color factor. We note that the particle multiplicity is shown here for illustrative purposes only as it is not used directly as input to \pnetee\ since it is a jet-based variable, while only particle-level observables are used. The remaining distributions of Figure~\ref{fig:input_comparison_idea_flavours} correspond to particle-level observables and are calculated using the charged constituent with the largest displacement. 
Figure~\ref{fig:input_comparison_idea_flavours} (b) displays the relative energy of the jet constituent with respect to the jet energy. Gluon jets populate lower values of this observables, indicating that the jet energy is more democratically distributed among the constituents. Figures~\ref{fig:input_comparison_idea_flavours} (b) display observables relevant for \emph{b} and \emph{c} quark identification, such as \sipIIDval\ (left) and its significance  \sipIIDsig\ (right) as defined in Table~\ref{tab:pnetinputs}. As expected, in \emph{b} jets, and to smaller extent in \emph{c} jets, a significantly larger displacement is observed compared to the other jet flavours. Displaced particles can also be present in other jet flavours, e.g. from long-lived $K_{\textrm{S}}^{0}$ or $\Lambda$ hadrons decays, but represent a much smaller fraction. 

\subsection{The flavour tagging algorithm}
\label{subsec:flavour_particlenet}
%Based on the motivations discussed in~\ref{sec:flavour}, we developed a new flavour tagging algorithm, \pnetee\, that aims to explore the full potential of the detectors and the superior FCC-ee environment. 
The \pnetee\ algorithm is based on the \pnet\ jet tagging algorithm~\cite{Qu:2019gqs}. \pnet\ uses an advanced network architecture, based on Graph Neural Networks (GNN) that first developed in the context of proton-proton collisions at the LHC. A novel jet representation was utilized in \pnet, where jets are represented as an un-ordered set of particles. As shown in Refs.~\cite{Qu:2019gqs,Mikuni:2020wpr,Mikuni:2021pou,Moreno:2019bmu,Moreno:2019neq,Bernreuther:2020vhm,Guo:2020vvt,Dreyer:2020brq,Konar:2021zdg,Dolan:2020qkr,Komiske:2018cqr}, this provides a more natural jet representation compared to alternative approaches based on jet images~\cite{Cogan:2014oua,Almeida:2015jua,deOliveira:2015xxd,Baldi:2016fql,Lin:2018cin,Barnard:2016qma,Komiske:2016rsd,Kasieczka:2017nvn,Macaluso:2018tck,Choi:2018dag} or ordered lists of jet constituents~\cite{Guest:2016iqz,Pearkes:2017hku,Egan:2017ojy,Fraser:2018ieu,Butter:2017cot,Kasieczka:2018lwf,Erdmann:2018shi,Louppe:2017ipp,Cheng:2017rdo} and translates to an improved tagging performance. A hierarchical learning approach using convolution operations~\cite{dgcnn} is adopted. Different convolutional layers are used to learn features at different scales: the shallower layers explore local neighborhood information, whereas more global structures are learned by deeper layers. The jet constituents are represented as a graph, where each node of the graph is a jet constituent, and relationships between the particles are the edges of the graph. Each node has a set of features related to constituent properties. However, the graph is not static, rather it is updated after each convolutional operation. The ultimate goal is to group jet constituents according to their proximity in the multi-dimensional space defined by the learned features. 

The current \pnetee\ implementation uses up to 75 constituents for each jet, sorted by the highest momentum,  which typically correspond to more than 99\% of the total jet momentum. The algorithm is designed to discriminate between five orthogonal jet classes:  \emph{ud},   \emph{s},  \emph{c},  \emph{b}, and  \emph{g}  jets. The training is performed using the \weaver\ package~\cite{weaver} on 10M jets (2M per category) over 30 epochs on a NVIDIA GTX 1080Ti GPUs. The network outputs 5 real numbers \ds{i}\ ($i = g,\,\ell(ud),\, s,\,c,\,b$) between 0 and 1 (discriminants), one for each jet category. Approximately 1M jets are used to evaluate the \pnetee\ performance. For every jet flavour pair $(i,j)$, the binary discriminant is constructed as:

\begin{equation}
\ds{i,j} = \frac{\ds{i}} {\ds{i} + \ds{j}},  
\end{equation}

where $\ds{i(j)}$ are the output scores of the classes $i$ and $j$. For example, $\ds{b,c}$ represents the binary discriminant for tagging \emph{b} quark jets against \emph{c} quark jets. The efficiency of tagging flavour $i$ as function of the probability of mis-identifiying the jet as flavour $j$ (mistag rate) can be constructed by computing the probability of selecting jets that satisfy $\ds{i,j} > \alpha$, for $\alpha \in [0,1]$. 
The receiver operating characteristic (ROC) curve, i.e. the mistag rate as a function of the tagging efficiency (for every $\alpha$), is used as a figure of merit for evaluating the tagger performance for every jet flavour.

\subsection{Results}
\label{subsec:results}

\begin{figure}[t]
     \centering 
     \subfigure[]{\includegraphics[width = 0.47\linewidth]{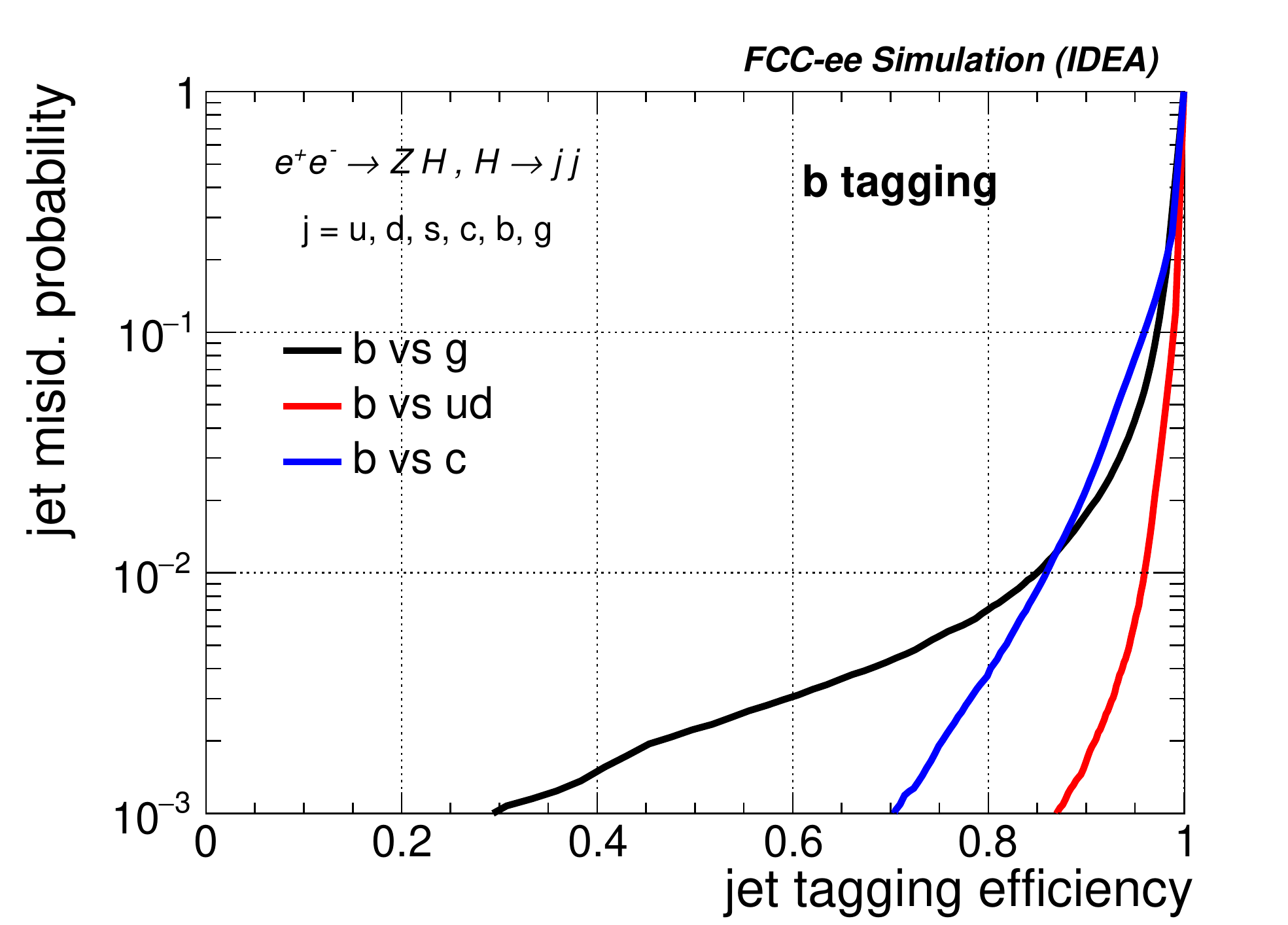}}
     \subfigure[]{\includegraphics[width = 0.47\linewidth]{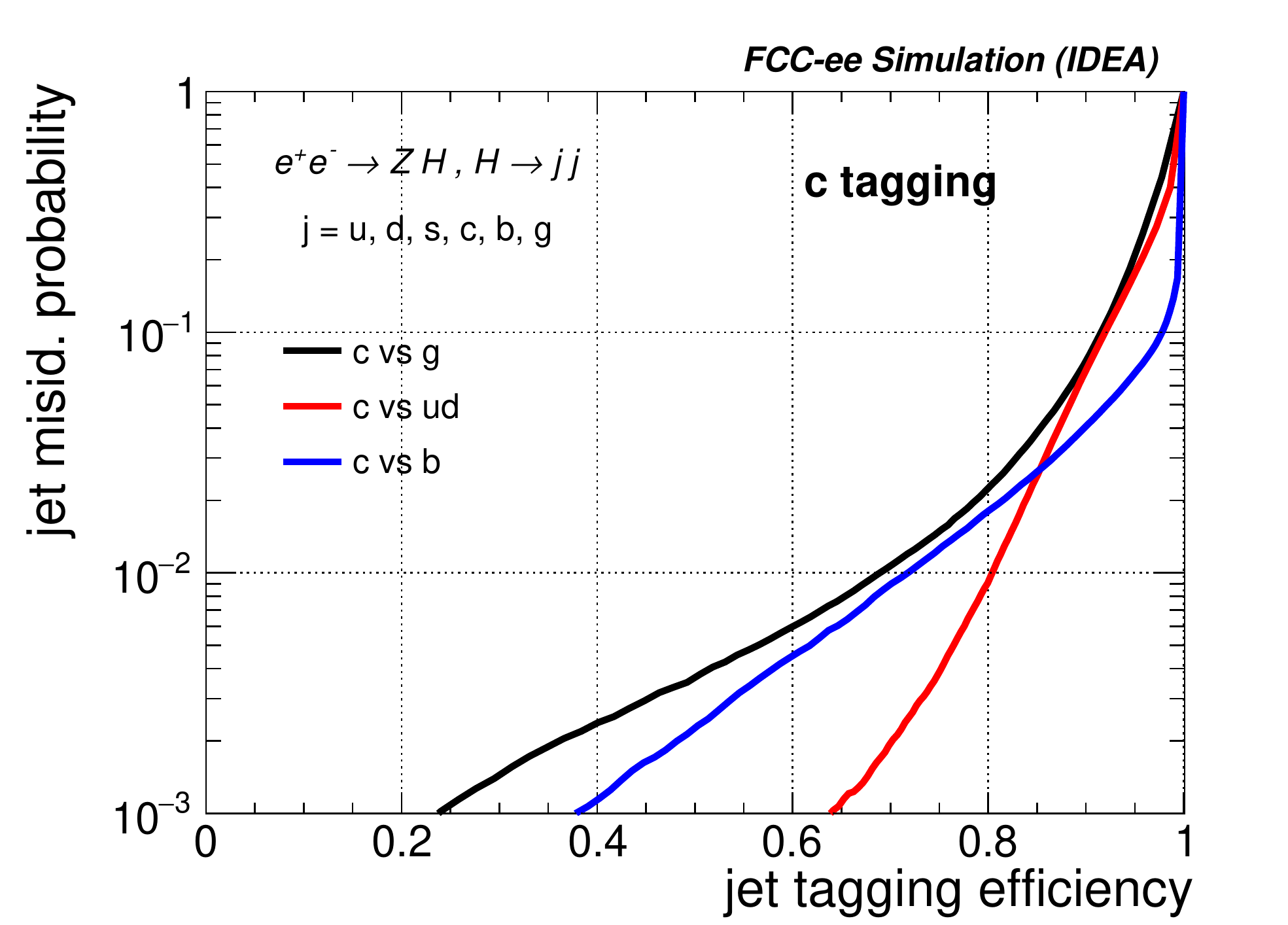}}
    
     \subfigure[]{\includegraphics[width = 0.47\linewidth]{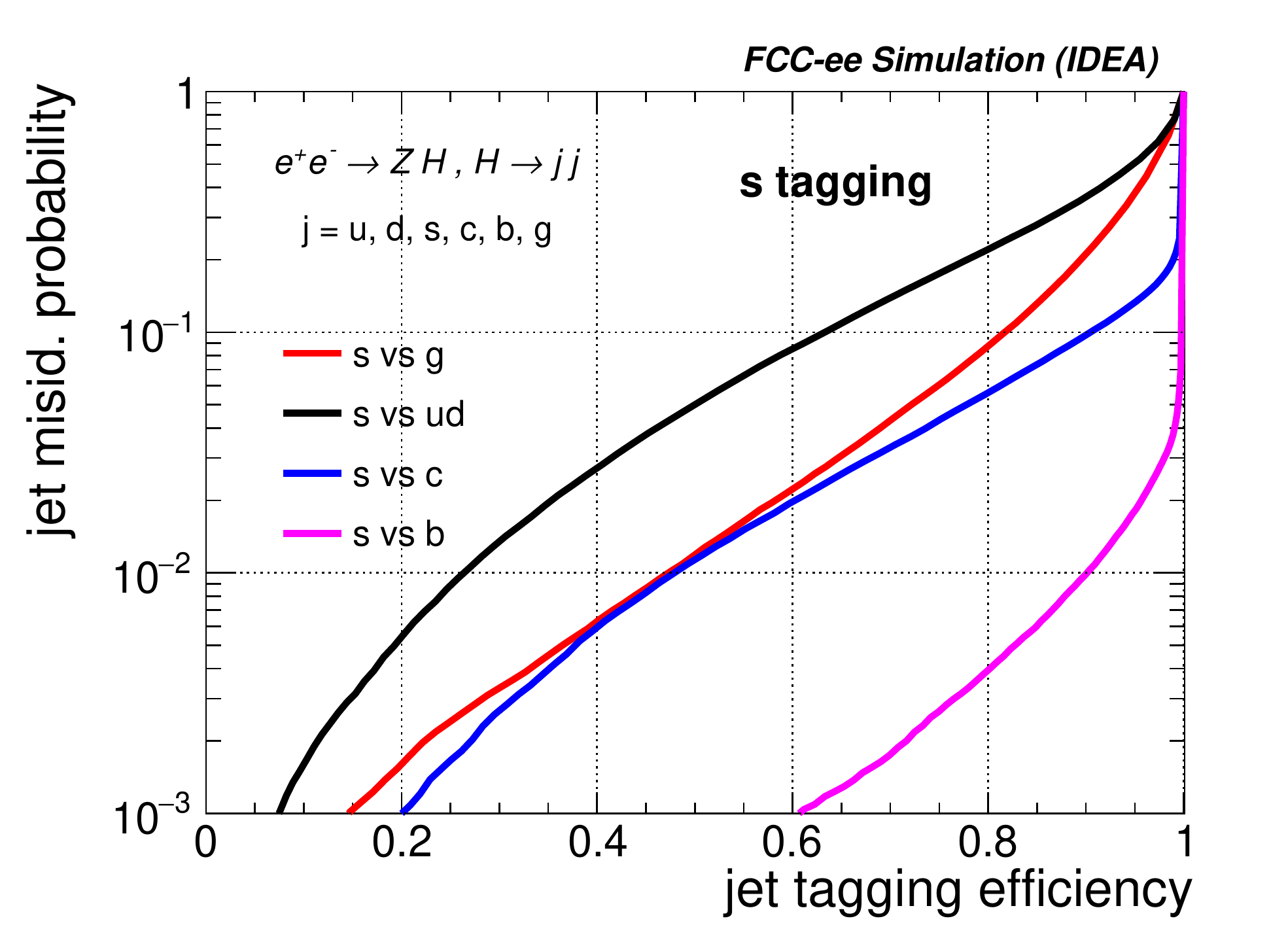}}
     \subfigure[]{\includegraphics[width = 0.47\linewidth]{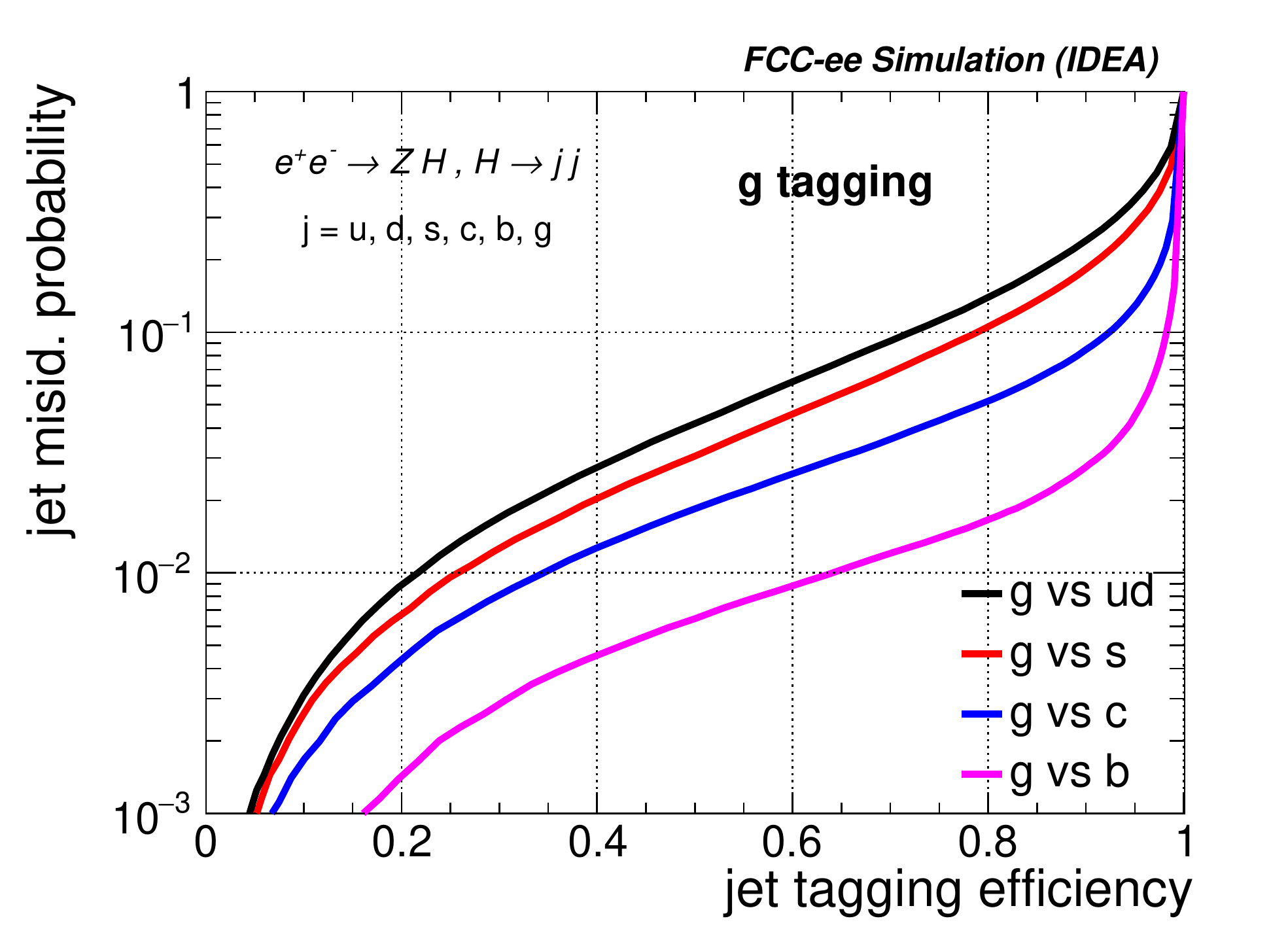}}
     
    \caption{\label{fig:pnetee_roc_default}Evaluation of \pnetee\ performance in terms of a receiver operating characteristic (ROC) curve for the identification of different jet flavours i.e.,  \emph{b}  quarks (upper left),  \emph{c}  quarks (upper right),  \emph{s}  (lower left), and  \emph{g}  (lower right). The different jet flavours considered background are indicated on the labels. The IDEA detector configuration is used.}
\end{figure}

The nominal \pnetee\ flavour tagging performance is shown in Figure~\ref{fig:pnetee_roc_default} for different jet flavours. The  \emph{b}  tagging performance is shown in Figure~\ref{fig:pnetee_roc_default}(a). The most effective discrimination is observed against  \emph{ud}  jets since these contain mostly tracks with no displacement. For high \emph{b} tagging efficiency,  \emph{g}  jet rejection is more effective than \emph{c} jet (with both being less effective than u, d or \emph{s} jet rejection). Conversely at small tagging efficiencies (i.e. for high tagging purity), \emph{c} jet rejection becomes more effective than  \emph{g}  jet rejection due to a sizeable probability for  \emph{g}  to produce \bbbar\ splittings. For \emph{c} tagging, at high efficiencies, \emph{b} jet discrimination is the most effective (due to a large difference of lifetime between B and D mesons), followed by \emph{ud} and  \emph{g}  jet rejection. For large \emph{c} tagging purity (i.e. at low efficiency and high background rejection), we observe that \emph{b} jet rejection becomes more challenging than \emph{ud} jet rejection, which is expected since a fraction of B meson have inherently a comparable decay length to D mesons. We also observe that in this regime \gcc\ splittings result into more challenging  \emph{g}  rejection. In Figure~\ref{fig:pnetee_roc_default}(c) the \emph{s} tagging performance is shown. The most effective discrimination is observed against \emph{b} jets followed by \emph{c} jets due to the large displacement of their tracks. The mistag rate against  \emph{g}  and \emph{ud} jets is substantially larger since displacement observables are not discriminating, and the algorithm relies mainly on PID-related variables. Rejection of \emph{g} jet is more effective than \emph{ud} jet one since \emph{s} and \emph{ud} jets have similar particle multiplicities. Finally, the  \emph{g}  tagging performance is displayed in Figure~\ref{fig:pnetee_roc_default}(d). Rejection of \emph{ud} jets is the most challenging, due to similar particle displacement and nature, followed by \emph{s}, \emph{c} and \emph{b} jet rejection. 

%$\epsilon_{\text{S}}$ and
%$\epsilon_{\text{B}}$, respectively, as a figure of merit. The efficiencies \esig~and
%\ebkg~are defined as:
%\begin{equation}
%\esig = \frac{N^{\text{tagged}}_\text{S}} {N^{\text{total}}_\text{S}} ~~~\text{and}~~~
%\ebkg = \frac{N^{\text{tagged}}_\text{B}} {N^{\text{total}}_\text{B}},
%\end{equation}
%where $N^{\text{tagged}}_{S}$ ($N^{\text{tagged}}_{B}$) is the number of
%signal (background) jets satisfying a selection on the \pnetee\ binary discriminant and $N^{\text{total}}_{S}$ ($N^{\text{total}}_{B}$) is
%the total number of generated particles considered to be signal (background).

%First, the \ebkg~as a function of \esig~is
%evaluated in terms of a receiver operating characteristic (ROC)
%curve in Figure~\ref{fig:pnetee_roc_default}, for different jet flavours as signal and background. The IDEA detector configuration is used.

\begin{figure}[t]
     \centering 
     \subfigure[]{\includegraphics[width = 0.47\linewidth]{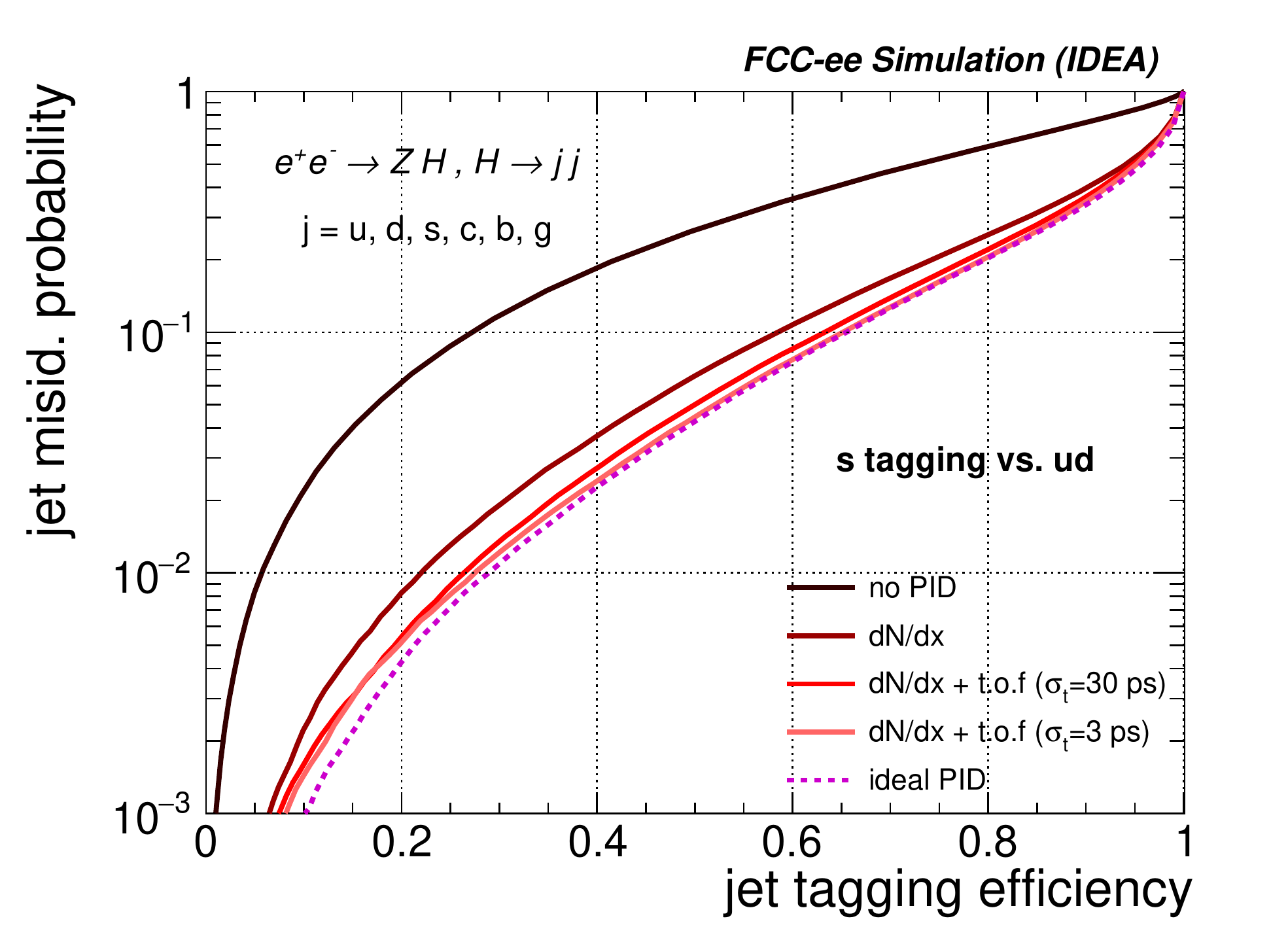}}
     \subfigure[]{\includegraphics[width = 0.47\linewidth]{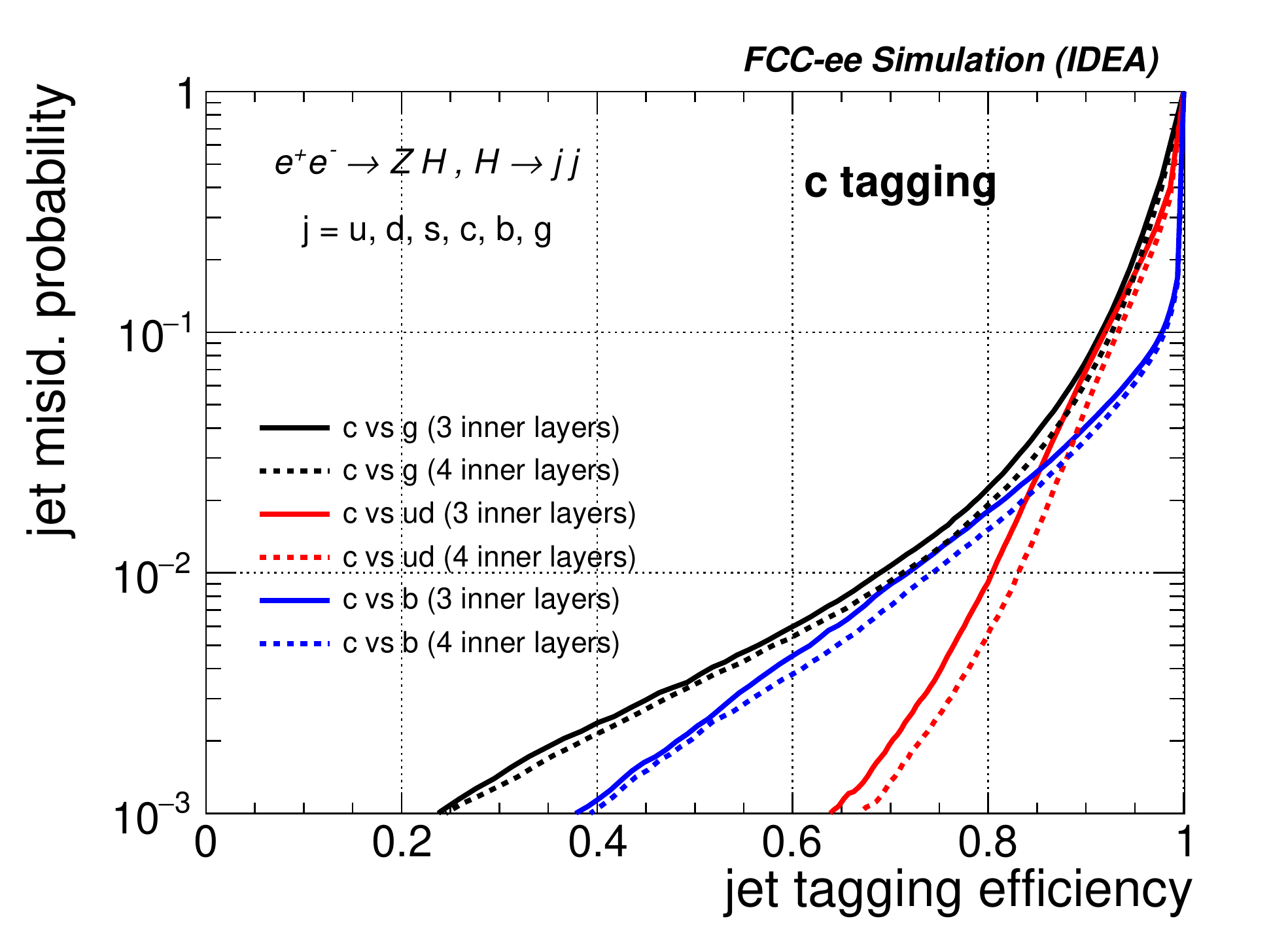}}
        \caption{\label{fig:pnetee_roc_detconf} Evaluation of \pnetee\ performance of the jet flavour identification for various detector assumptions. (a) Impact of particle identification on \emph{s} tagging performance. (b) Impact of inner track geometry 3 vs 4 layers on \emph{c} tagging.}
\end{figure}

The modularity of the framework enables the study of the algorithm performance for different detector design choices. In this work, we report two representative examples of possible detector design variations. Figure~\ref{fig:pnetee_roc_detconf} (a) shows the importance of particle identification information in discriminating  \emph{s}  jets from other jet flavours. Exploiting PID information with the nominal \dndx\ and \tof\ resolutions yields to approximately an order of magnitude reduced \emph{ud} jet mistag rate for the same \emph{s}  tagging efficiency. A timing detector providing an improved \tof\ resolution of 3~ps for charged particles, yields a small, but detectable improvement compared to the more realistic scenario of 30~ps. The performance obtained using MC truth information for PID ("ideal PID") is also shown for reference. In that case only a marginal improvement in performance is observed, suggesting that the existing detector configurations and PID algorithms are very close to optimal. The usage of PID information brings only modest improvement in other jet flavour tasks.

The distance of the first vertex detector layer to the interaction point is the most important parameter for achieving optimal transverse impact parameter resolution and hence \emph{b} and \emph{c} tagging performance. While the nominal IDEA vertex detector provides already an excellent resolution (three layers, with the innermost layer located at 1.5~cm), we study the impact of introducing an additional fourth layer in the pixel detector, closer to the beam pipe, located at 1 cm from the interaction point, on \emph{c} jet identification. The corresponding performance is displayed in Figure~\ref{fig:pnetee_roc_detconf} (b). The largest improvement is observed in the discrimination against \emph{ud} jets, where for the same \esig, \ebkg\ is reduced by almost a factor of two. Smaller, yet important improvement, is seen in the discrimination against other jet flavours. The impact of an additional pixel layer was studied for other jet flavours treated as signal without significant improvement in the performance. 

%%%%%%%%%%%%%%%%%%%%%%%%%%%%%%%%%%%%%%%%%%%%%%%%%%%%%

\section{Conclusion and perspectives}
\label{sec:disc}

Jet flavour tagging will be a crucial tool for maximising the physics potential at future colliders. This work builds on the design of a fast detector simulation framework, and provides an efficient way to study the impact of different detector design options to the jet flavour tagging problem. A fast tracking module was developed, which allows to easily configure a full tracking geometry including material effects and compute both the charge particle track parameters and the track covariance matrix. Two algorithms that allow for particle identification, the time-of-flight and cluster counting with respectively configurable time resolution and gas composition have also been added. The framework is designed to provide flexibility for further studies, such as the exploration of alternative clustering algorithms, beam energies and final states.

Deep learning techniques based on GNNs have proven very effective for classification problems such as jet flavour tagging and boosted jet tagging at the LHC, and have not been explored yet in the context of future experiments. This paper presents the first algorithm for jet flavour tagging at future \ee\ colliders using state-of-the-art jet representation and a GNN architecture. At such future machines where statistics for Higgs processes are moderate, flavour taggers will be required to perform well in the high tagging efficiency regime while still providing excellent background rejection. In this study we have investigated the impact of MIP timing resolution and of an additional inner tracking layer on the tagging performance. More studies are possible and should be pursued: the interplay of MIP and calorimeter timing on PID performance, the impact of the tracker design on displaced tracks performance, \ks\ and $\Lambda$ reconstruction and hence on  \emph{s}  tagging, secondary vertex reconstruction on  \emph{b},  \emph{c}  and  \emph{s} tagging. Another area for future studies is the calibration of the algorithm. The algorithm is designed to have very little dependence on the jet kinematics and therefore a calibration strategy relying on a Z boson sample of unprecedented statistical power expected to be obtained in \ee\ experiments seems a promising avenue. 

We stress that this study has possible limitations given the inherent optimistic nature of fast simulation. In particular, this tracking simulation include a simplistic particle-matter description where multiple scattering is taken into account to derive track parameter resolutions but no secondary emissions are simulated (i.e. electron bremsstrahlung and hadronic interactions are neglected). A natural next step is to assess the limitation of the fast detector simulation framework by validating the results with events produced using Full Simulation. Nevertheless, the set of tools presented in this article should provide robust means for assessing an upper limit of the achievable tagging performance and the relative performance of alternative detector design choices at future \ee\ colliders. We also point out that the presented framework should allow for similar optimisations at any future machine, including high energy proton-proton or Muon colliders, acknowledging however that further caution is required due to the lack of simulation of larger background levels. 

%\begin{table}[ht!]
%\renewcommand{\arraystretch}{1.5}
%\begin{center}
%\begin{tabular}{lcccc}
%Process & $\sigma$(14 TeV) & $\sigma$(100 TeV) & accuracy & K-factor\\ \hline
%\gghh\ & $36.69 \pm 5.3\%$ & $1224 \pm 5.6\%$ & $\mathrm{\nnlo_{FTapprox}}$ & 1.08 \\ 
%\vbfhh\ & $2.05 \pm 2.1\%$ & $82.8 \pm 2.1\%$ & N$^3$LO & 1.15\\ 
%\tthh\ & $0.949 \pm 2.9\%$ & $82.1 \pm 7.8\%$ & \nlo & 1.38\\ 
%\vhh\ & $0.982 \pm 1.8\%$ & $16.23 \pm  2.9 \%$ & \nnlo & 1.40\\
%\end{tabular} %}
%\caption{a caption}
%\label{tab:xsecs_s}
%\end{center}
%\end{table}

%\begin{figure}[t!]
%\centering     %%% not \center
%  \subfigure[]
%  {\label{fig:xsec_vs_lambda}
%   \includegraphics[width = %0.47\linewidth]{Fig/xsec_vs_klambda.pdf}
%  }
%  \subfigure[]
%  {\label{fig:hhptgen}
%   \includegraphics[width = 0.47\linewidth]{Fig/pwp8_hhpt.pdf}
%  }
%  \caption{(a) a figure with subfigures}
%\end{figure}

%%%%%%%%%%%%%%%%%%%%%%%%%%%%%%%%%%%%%%%%%%%%%%%%%%%%%

\begin{acknowledgments}

We acknowledge the support from CERN and INFN for this work. In particular, we would like to thank our colleagues from the FCC physics, experiments and detectors group (PED). In particular we thank Patrizia Azzi, Alain Blondel, Patrick Janot, Emmanuel Perez and Gavin Salam for helpful discussions and suggestions. We are also grateful to Sylvie Braibant for the extensive testing of the \trackcovMod\ module in \delphes\ and to Pavel Demin and Huilin Qu for the valuable help and support on the \delphes\ and \weaver\ frameworks respectively. 

\end{acknowledgments}

\bibliographystyle{JHEP}
\clearpage
\bibliography{paper}

\providecommand{\href}[2]{#2}\begingroup\raggedright\begin{thebibliography}{10}

\bibitem{Abada:2019zxq}
{\scshape FCC} collaboration, A.~Abada et~al., \emph{{FCC-ee: The Lepton
  Collider}: {Future Circular Collider Conceptual Design Report Volume 2}},
  \href{https://doi.org/10.1140/epjst/e2019-900045-4}{\emph{Eur. Phys. J. ST}
  {\bfseries 228} (2019) 261}.

\bibitem{CEPCStudyGroup:2018ghi}
{\scshape CEPC Study Group} collaboration, M.~Dong et~al., \emph{{CEPC
  Conceptual Design Report: Volume 2 - Physics \& Detector}},
  \href{https://arxiv.org/abs/1811.10545}{{\ttfamily arXiv:1811.10545}}
  [hep-ex].

\bibitem{Baer:2013cma}
\emph{{The International Linear Collider Technical Design Report - Volume 2:
  Physics}},  \href{https://arxiv.org/abs/1306.6352}{{\ttfamily
  arXiv:1306.6352}} [hep-ph].

\bibitem{LC-REP-2013-021}
J.~Tian,  and K.~Fujii, \emph{{Summary of Higgs coupling measurements with
  staged running of ILC at 250 GeV, 500 GeV and 1 TeV}},  Tech. Rep.
  LC-REP-2013-021, DESY, Jul, 2013.

\bibitem{CLICdp:2018cto}
{\scshape CLICdp, CLIC} collaboration, T.~K. Charles et~al., \emph{{The Compact
  Linear Collider (CLIC) - 2018 Summary Report}},
  \href{https://arxiv.org/abs/1812.06018}{{\ttfamily arXiv:1812.06018}}
  [physics.acc-ph].

\bibitem{Benedikt:2651300}
M.~Benedikt, M.~Capeans~Garrido, F.~Cerutti, B.~Goddard, J.~Gutleber, J.~M.
  Jimenez et~al., \emph{{Future Circular Collider Study. Volume 3: The Hadron
  Collider (FCC-hh)}},  Tech. Rep. CERN-ACC-2018-0058, CERN, Geneva, Dec, 2018,
  \href{https://cds.cern.ch/record/2651300}{https://cds.cern.ch/record/2651300}.

\bibitem{Asner:2013psa}
D.~M. Asner et~al., \emph{{ILC Higgs White Paper}},  in \emph{{Community Summer
  Study 2013}: {Snowmass on the Mississippi}}, 10, 2013,
  \href{https://arxiv.org/abs/1310.0763}{{\ttfamily arXiv:1310.0763}} [hep-ph].

\bibitem{Thomson:2015jda}
M.~Thomson, \emph{{Model-independent measurement of the e$^{{+}}$ e$^{-}$
  $\rightarrow $ HZ cross section at a future e$^{{+}}$ e$^{-}$ linear collider
  using hadronic Z decays}},
  \href{https://doi.org/10.1140/epjc/s10052-016-3911-5}{\emph{Eur. Phys. J. C}
  {\bfseries 76} (2016) 72} \href{https://arxiv.org/abs/1509.02853}{{\ttfamily
  arXiv:1509.02853}} [hep-ex].

\bibitem{Abramowicz:2016zbo}
H.~Abramowicz et~al., \emph{{Higgs physics at the CLIC
  electron\textendash{}positron linear collider}},
  \href{https://doi.org/10.1140/epjc/s10052-017-4968-5}{\emph{Eur. Phys. J. C}
  {\bfseries 77} (2017) 475} \href{https://arxiv.org/abs/1608.07538}{{\ttfamily
  arXiv:1608.07538}} [hep-ex].

\bibitem{deBlas:2019rxi}
J.~de~Blas et~al., \emph{{Higgs Boson Studies at Future Particle Colliders}},
  \href{https://doi.org/10.1007/JHEP01(2020)139}{\emph{JHEP} {\bfseries 01}
  (2020) 139} \href{https://arxiv.org/abs/1905.03764}{{\ttfamily
  arXiv:1905.03764}} [hep-ph].

\bibitem{An:2018dwb}
F.~An et~al., \emph{{Precision Higgs physics at the CEPC}},
  \href{https://doi.org/10.1088/1674-1137/43/4/043002}{\emph{Chin. Phys. C}
  {\bfseries 43} (2019) 043002}
  \href{https://arxiv.org/abs/1810.09037}{{\ttfamily arXiv:1810.09037}}
  [hep-ex].

\bibitem{L.Borgonovi:2642471}
L.~Borgonovi, S.~Braibant, B.~Di~Micco, E.~Fontanesi, P.~Harris, C.~Helsens
  et~al., \emph{{Higgs measurements at FCC-hh}},  Tech. Rep.
  CERN-ACC-2018-0045, CERN, Geneva, Oct, 2018,
  \href{https://cds.cern.ch/record/2642471}{https://cds.cern.ch/record/2642471}.

\bibitem{Koratzinos:2013ncw}
M.~Koratzinos et~al., \emph{{TLEP: A High-Performance Circular $e^+e^-$
  Collider to Study the Higgs Boson}},  in \emph{{4th International Particle
  Accelerator Conference}}, p.~TUPME040, 2013,
  \href{https://arxiv.org/abs/1305.6498}{{\ttfamily arXiv:1305.6498}}
  [physics.acc-ph].

\bibitem{Mangano:2020sao}
M.~L. Mangano, G.~Ortona and M.~Selvaggi, \emph{{Measuring the Higgs
  self-coupling via Higgs-pair production at a 100 TeV p-p collider}},
  \href{https://doi.org/10.1140/epjc/s10052-020-08595-3}{\emph{Eur. Phys. J. C}
  {\bfseries 80} (2020) 1030}
  \href{https://arxiv.org/abs/2004.03505}{{\ttfamily arXiv:2004.03505}}
  [hep-ph].

\bibitem{Seidel:2013sqa}
K.~Seidel, F.~Simon, M.~Tesar and S.~Poss, \emph{{Top quark mass measurements
  at and above threshold at CLIC}},
  \href{https://doi.org/10.1140/epjc/s10052-013-2530-7}{\emph{Eur. Phys. J. C}
  {\bfseries 73} (2013) 2530} \href{https://arxiv.org/abs/1303.3758}{{\ttfamily
  arXiv:1303.3758}} [hep-ex].

\bibitem{Janot:2015yza}
P.~Janot, \emph{{Top-quark electroweak couplings at the FCC-ee}},
  \href{https://doi.org/10.1007/JHEP04(2015)182}{\emph{JHEP} {\bfseries 04}
  (2015) 182} \href{https://arxiv.org/abs/1503.01325}{{\ttfamily
  arXiv:1503.01325}} [hep-ph].

\bibitem{Mangano:2015aow}
M.~L. Mangano, T.~Plehn, P.~Reimitz, T.~Schell and H.-S. Shao, \emph{{Measuring
  the Top Yukawa Coupling at 100 TeV}},
  \href{https://doi.org/10.1088/0954-3899/43/3/035001}{\emph{J. Phys. G}
  {\bfseries 43} (2016) 035001}
  \href{https://arxiv.org/abs/1507.08169}{{\ttfamily arXiv:1507.08169}}
  [hep-ph].

\bibitem{Azzi:2021gwg}
P.~Azzi, L.~Gouskos, M.~Selvaggi and F.~Simon, \emph{{Higgs and top physics
  reconstruction challenges and opportunities at FCC-ee}},
  \href{https://doi.org/10.1140/epjp/s13360-021-02223-z}{\emph{Eur. Phys. J.
  Plus} {\bfseries 137} (2022) 39}
  \href{https://arxiv.org/abs/2107.05003}{{\ttfamily arXiv:2107.05003}}
  [hep-ex].

\bibitem{Abdallah:2002xm}
{\scshape DELPHI} collaboration, J.~Abdallah et~al., \emph{{b tagging in DELPHI
  at LEP}}, \href{https://doi.org/10.1140/epjc/s2003-01441-8}{\emph{Eur. Phys.
  J. C} {\bfseries 32} (2004) 185}
  \href{https://arxiv.org/abs/hep-ex/0311003}{{\ttfamily
  arXiv:hep-ex/0311003}}.

\bibitem{Proriol:1950599}
J.~Proriol, A.~Falvard, P.~Henrard, J.~Jousset and B.~B., \emph{{Tagging B
  quark events in ALEPH with neural networks: comparison of different
  methods}}, {\emph{Int. J. Neural Syst.} {\bfseries 3 Supp.} (1991) 267}.

\bibitem{D0:2010zho}
{\scshape D0} collaboration, V.~M. Abazov et~al., \emph{{$b$-Jet Identification
  in the D0 Experiment}},
  \href{https://doi.org/10.1016/j.nima.2010.03.118}{\emph{Nucl. Instrum. Meth.
  A} {\bfseries 620} (2010) 490}
  \href{https://arxiv.org/abs/1002.4224}{{\ttfamily arXiv:1002.4224}} [hep-ex].

\bibitem{Freeman:2012uf}
J.~Freeman, T.~Junk, M.~Kirby, Y.~Oksuzian, T.~J. Phillips, F.~D. Snider
  et~al., \emph{{Introduction to HOBIT, a b-Jet Identification Tagger at the
  CDF Experiment Optimized for Light Higgs Boson Searches}},
  \href{https://doi.org/10.1016/j.nima.2012.09.021}{\emph{Nucl. Instrum. Meth.
  A} {\bfseries 697} (2013) 64}
  \href{https://arxiv.org/abs/1205.1812}{{\ttfamily arXiv:1205.1812}} [hep-ex].

\bibitem{ATLAS-CONF-2010-042}
{\scshape ATLAS} collaboration, \emph{{Performance of the ATLAS Secondary
  Vertex b-tagging Algorithm in 7 TeV Collision Data}},  Tech. Rep.
  ATLAS-CONF-2010-042, CERN, Geneva, Jul, 2010,
  \href{https://cds.cern.ch/record/1277682}{https://cds.cern.ch/record/1277682}.

\bibitem{ATLAS-CONF-2010-070}
{\scshape ATLAS} collaboration, \emph{{Tracking Studies for $b$-tagging with 7
  TeV Collision Data with the ATLAS Detector}},  Tech. Rep.
  ATLAS-CONF-2010-070, CERN, Geneva, Jul, 2010,
  \href{https://cds.cern.ch/record/1281352}{https://cds.cern.ch/record/1281352}.

\bibitem{ATLAS-CONF-2010-091}
{\scshape ATLAS} collaboration, \emph{{Performance of Impact Parameter-Based
  b-tagging Algorithms with the ATLAS Detector using Proton-Proton Collisions
  at $\sqrt{s}$ = 7 TeV}},  Tech. Rep. ATLAS-CONF-2010-091, CERN, Geneva, Oct,
  2010,
  \href{https://cds.cern.ch/record/1299106}{https://cds.cern.ch/record/1299106}.

\bibitem{CMS-PAS-BTV-11-004}
{\scshape CMS} collaboration, \emph{{b-Jet Identification in the CMS
  Experiment}},  Tech. Rep. CMS-PAS-BTV-11-004, CERN, Geneva, 2012,
  \href{http://cds.cern.ch/record/1427247}{http://cds.cern.ch/record/1427247}.

\bibitem{ATLASBTag2016}
{\scshape ATLAS} collaboration, \emph{Performance of b-jet identification in
  the {ATLAS} experiment},
  \href{https://doi.org/10.1088/1748-0221/11/04/p04008}{\emph{JINST} {\bfseries
  11} (2016) P04008}.

\bibitem{CMS:JME}
{\scshape CMS} collaboration, \emph{Identification of heavy, energetic,
  hadronically decaying particles using machine-learning techniques},
  \href{https://doi.org/10.1088/1748-0221/15/06/p06005}{\emph{Journal of
  Instrumentation} {\bfseries 15} (2020) P06005}
  \href{https://arxiv.org/abs/arXiv:2004.08262}{{\ttfamily
  arXiv:arXiv:2004.08262}}.

\bibitem{ATL-PHYS-PUB-2017-003}
{ATLAS collaboration}, \emph{{Identification of Jets Containing $b$-Hadrons
  with Recurrent Neural Networks at the ATLAS Experiment}},  Tech. Rep.
  ATL-PHYS-PUB-2017-003, CERN, Geneva, 2017,
  \href{https://cds.cern.ch/record/2255226}{https://cds.cern.ch/record/2255226}.

\bibitem{Bols:2020bkb}
E.~Bols, J.~Kieseler, M.~Verzetti, M.~Stoye and A.~Stakia, \emph{{Jet Flavour
  Classification Using DeepJet}},
  \href{https://doi.org/10.1088/1748-0221/15/12/P12012}{\emph{JINST} {\bfseries
  15} (2020) P12012} \href{https://arxiv.org/abs/2008.10519}{{\ttfamily
  arXiv:2008.10519}} [hep-ex].

\bibitem{CMS:2017yfk}
{\scshape CMS} collaboration, A.~M. Sirunyan et~al., \emph{{Particle-flow
  reconstruction and global event description with the CMS detector}},
  \href{https://doi.org/10.1088/1748-0221/12/10/P10003}{\emph{JINST} {\bfseries
  12} (2017) P10003} \href{https://arxiv.org/abs/1706.04965}{{\ttfamily
  arXiv:1706.04965}} [physics.ins-det].

\bibitem{tc_module}
``{\trackcovMod\ module in \delphes}.''
  \url{https://github.com/delphes/delphes/blob/master/modules/TrackCovariance.cc}.

\bibitem{tof_module}
``{\tofMod\ module in \delphes}.''
  \url{https://github.com/delphes/delphes/blob/master/modules/TimeOfFlight.cc}.

\bibitem{cc_module}
``{\ccMod\ module in \delphes}.''
  \url{https://github.com/delphes/delphes/blob/master/modules/ClusterCounting.cc}.

\bibitem{Qu:2019gqs}
H.~Qu and L.~Gouskos, \emph{{ParticleNet: Jet Tagging via Particle Clouds}},
  \href{https://doi.org/10.1103/PhysRevD.101.056019}{\emph{Phys. Rev. D}
  {\bfseries 101} (2020) 056019}
  \href{https://arxiv.org/abs/1902.08570}{{\ttfamily arXiv:1902.08570}}
  [hep-ph].

\bibitem{Bedeschi:2021nln}
F.~Bedeschi, \emph{{A detector concept proposal for a circular $e^+e^-$
  collider}}, \href{https://doi.org/10.22323/1.390.0819}{\emph{PoS} {\bfseries
  ICHEP2020} (2021) 819}.

\bibitem{delphes_card_idea}
``{FCC-ee IDEA detector \delphes\ card}.''
  \url{https://github.com/delphes/delphes/blob/master/cards/delphes_card_IDEA.tcl}.

\bibitem{Drasal:2018zij}
Z.~Drasal and W.~Riegler, \emph{{An extension of the Gluckstern formulae for
  multiple scattering: Analytic expressions for track parameter resolution
  using optimum weights}},
  \href{https://doi.org/10.1016/j.nima.2018.08.078}{\emph{Nucl. Instrum. Meth.
  A} {\bfseries 910} (2018) 127}
  \href{https://arxiv.org/abs/1805.12014}{{\ttfamily arXiv:1805.12014}}
  [physics.ins-det].

\bibitem{Walenta:1979ut}
A.~H. Walenta, \emph{{The time expansion chamber and single ionization cluster
  measurment.}}, \href{https://doi.org/10.1109/TNS.1979.4329616}{\emph{IEEE
  Trans. Nucl. Sci.} {\bfseries 26} (1979) 73}.

\bibitem{Caron:2013gca}
J.-F. Caron et~al., \emph{{Improved Particle Identification Using Cluster
  Counting in a Full-Length Drift Chamber Prototype}},
  \href{https://doi.org/10.1016/j.nima.2013.09.028}{\emph{Nucl. Instrum. Meth.
  A} {\bfseries 735} (2014) 169}
  \href{https://arxiv.org/abs/1307.8101}{{\ttfamily arXiv:1307.8101}}
  [physics.ins-det].

\bibitem{Smirnov:2005yi}
I.~B. Smirnov, \emph{{Modeling of ionization produced by fast charged particles
  in gases}}, \href{https://doi.org/10.1016/j.nima.2005.08.064}{\emph{Nucl.
  Instrum. Meth. A} {\bfseries 554} (2005) 474}.

\bibitem{Veenhof:1998tt}
R.~Veenhof, \emph{{GARFIELD, recent developments}},
  \href{https://doi.org/10.1016/S0168-9002(98)00851-1}{\emph{Nucl. Instrum.
  Meth. A} {\bfseries 419} (1998) 726}.

\bibitem{Alwall:2014hca}
J.~Alwall, R.~Frederix, S.~Frixione, V.~Hirschi, F.~Maltoni, O.~Mattelaer
  et~al., \emph{{The automated computation of tree-level and next-to-leading
  order differential cross sections, and their matching to parton shower
  simulations}}, \href{https://doi.org/10.1007/JHEP07(2014)079}{\emph{JHEP}
  {\bfseries 07} (2014) 079} \href{https://arxiv.org/abs/1405.0301}{{\ttfamily
  arXiv:1405.0301}} [hep-ph].

\bibitem{Sjostrand:2014zea}
T.~Sj{\"o}strand, S.~Ask, J.~R. Christiansen, R.~Corke, N.~Desai, P.~Ilten
  et~al., \emph{{An Introduction to PYTHIA 8.2}},
  \href{https://doi.org/10.1016/j.cpc.2015.01.024}{\emph{Comput. Phys. Commun.}
  {\bfseries 191} (2015) 159} \href{https://arxiv.org/abs/1410.3012}{{\ttfamily
  arXiv:1410.3012}} [hep-ph].

\bibitem{dr_module}
``{\drMod\ module in \delphes}.''
  \url{https://github.com/delphes/delphes/blob/master/modules/DualReadoutCalorimeter.cc}.

\bibitem{Cacciari:2011ma}
M.~Cacciari, G.~P. Salam and G.~Soyez, \emph{{FastJet User Manual}},
  \href{https://doi.org/10.1140/epjc/s10052-012-1896-2}{\emph{Eur. Phys. J.}
  {\bfseries C72} (2012) 1896}
  \href{https://arxiv.org/abs/1111.6097}{{\ttfamily arXiv:1111.6097}} [hep-ph].

\bibitem{Cacciari:2008gp}
M.~Cacciari, G.~P. Salam and G.~Soyez, \emph{{The anti-$k_t$ jet clustering
  algorithm}}, \href{https://doi.org/10.1088/1126-6708/2008/04/063}{\emph{JHEP}
  {\bfseries 04} (2008) 063} \href{https://arxiv.org/abs/0802.1189}{{\ttfamily
  arXiv:0802.1189}} [hep-ph].

\bibitem{CATANI1991432}
S.~Catani, Y.~Dokshitzer, M.~Olsson, G.~Turnock and B.~Webber, \emph{New
  clustering algorithm for multijet cross sections in \ee\ annihilation},
  \href{https://doi.org/https://doi.org/10.1016/0370-2693(91)90196-W}{\emph{Physics
  Letters B} {\bfseries 269} (1991) 432}.

\bibitem{Mikuni:2020wpr}
V.~Mikuni and F.~Canelli, \emph{{ABCNet: An attention-based method for particle
  tagging}}, \href{https://doi.org/10.1140/epjp/s13360-020-00497-3}{\emph{Eur.
  Phys. J. Plus} {\bfseries 135} (2020) 463}
  \href{https://arxiv.org/abs/2001.05311}{{\ttfamily arXiv:2001.05311}}
  [physics.data-an].

\bibitem{Mikuni:2021pou}
V.~Mikuni and F.~Canelli, \emph{{Point cloud transformers applied to collider
  physics}}, \href{https://doi.org/10.1088/2632-2153/ac07f6}{\emph{Mach. Learn.
  Sci. Tech.} {\bfseries 2} (2021) 035027}
  \href{https://arxiv.org/abs/2102.05073}{{\ttfamily arXiv:2102.05073}}
  [physics.data-an].

\bibitem{Moreno:2019bmu}
E.~A. Moreno, O.~Cerri, J.~M. Duarte, H.~B. Newman, T.~Q. Nguyen, A.~Periwal
  et~al., \emph{{JEDI-net: a jet identification algorithm based on interaction
  networks}}, \href{https://doi.org/10.1140/epjc/s10052-020-7608-4}{\emph{Eur.
  Phys. J. C} {\bfseries 80} (2020) 58}
  \href{https://arxiv.org/abs/1908.05318}{{\ttfamily arXiv:1908.05318}}
  [hep-ex].

\bibitem{Moreno:2019neq}
E.~A. Moreno, T.~Q. Nguyen, J.-R. Vlimant, O.~Cerri, H.~B. Newman, A.~Periwal
  et~al., \emph{{Interaction networks for the identification of boosted $H
  \rightarrow b\overline{b}$ decays}},
  \href{https://doi.org/10.1103/PhysRevD.102.012010}{\emph{Phys. Rev. D}
  {\bfseries 102} (2020) 012010}
  \href{https://arxiv.org/abs/1909.12285}{{\ttfamily arXiv:1909.12285}}
  [hep-ex].

\bibitem{Bernreuther:2020vhm}
E.~Bernreuther, T.~Finke, F.~Kahlhoefer, M.~Kr\"amer and A.~M\"uck,
  \emph{{Casting a graph net to catch dark showers}},
  \href{https://doi.org/10.21468/SciPostPhys.10.2.046}{\emph{SciPost Phys.}
  {\bfseries 10} (2021) 046} \href{https://arxiv.org/abs/2006.08639}{{\ttfamily
  arXiv:2006.08639}} [hep-ph].

\bibitem{Guo:2020vvt}
J.~Guo, J.~Li, T.~Li and R.~Zhang, \emph{{Boosted Higgs boson jet
  reconstruction via a graph neural network}},
  \href{https://doi.org/10.1103/PhysRevD.103.116025}{\emph{Phys. Rev. D}
  {\bfseries 103} (2021) 116025}
  \href{https://arxiv.org/abs/2010.05464}{{\ttfamily arXiv:2010.05464}}
  [hep-ph].

\bibitem{Dreyer:2020brq}
F.~A. Dreyer and H.~Qu, \emph{{Jet tagging in the Lund plane with graph
  networks}}, \href{https://doi.org/10.1007/JHEP03(2021)052}{\emph{JHEP}
  {\bfseries 03} (2021) 052} \href{https://arxiv.org/abs/2012.08526}{{\ttfamily
  arXiv:2012.08526}} [hep-ph].

\bibitem{Konar:2021zdg}
P.~Konar, V.~S. Ngairangbam and M.~Spannowsky, \emph{{Energy-weighted Message
  Passing: an infra-red and collinear safe graph neural network algorithm}},
  \href{https://arxiv.org/abs/2109.14636}{{\ttfamily arXiv:2109.14636}}
  [hep-ph].

\bibitem{Dolan:2020qkr}
M.~J. Dolan and A.~Ore, \emph{{Equivariant Energy Flow Networks for Jet
  Tagging}}, \href{https://doi.org/10.1103/PhysRevD.103.074022}{\emph{Phys.
  Rev. D} {\bfseries 103} (2021) 074022}
  \href{https://arxiv.org/abs/2012.00964}{{\ttfamily arXiv:2012.00964}}
  [hep-ph].

\bibitem{Komiske:2018cqr}
P.~T. Komiske, E.~M. Metodiev and J.~Thaler, \emph{{Energy Flow Networks: Deep
  Sets for Particle Jets}},
  \href{https://doi.org/10.1007/JHEP01(2019)121}{\emph{JHEP} {\bfseries 01}
  (2019) 121} \href{https://arxiv.org/abs/1810.05165}{{\ttfamily
  arXiv:1810.05165}} [hep-ph].

\bibitem{Cogan:2014oua}
J.~Cogan, M.~Kagan, E.~Strauss and A.~Schwarztman, \emph{{Jet-Images: Computer
  Vision Inspired Techniques for Jet Tagging}},
  \href{https://doi.org/10.1007/JHEP02(2015)118}{\emph{JHEP} {\bfseries 02}
  (2015) 118} \href{https://arxiv.org/abs/1407.5675}{{\ttfamily
  arXiv:1407.5675}} [hep-ph].

\bibitem{Almeida:2015jua}
L.~G. Almeida, M.~Backovi\'c, M.~Cliche, S.~J. Lee and M.~Perelstein,
  \emph{{Playing Tag with ANN: Boosted Top Identification with Pattern
  Recognition}}, \href{https://doi.org/10.1007/JHEP07(2015)086}{\emph{JHEP}
  {\bfseries 07} (2015) 086} \href{https://arxiv.org/abs/1501.05968}{{\ttfamily
  arXiv:1501.05968}} [hep-ph].

\bibitem{deOliveira:2015xxd}
L.~de~Oliveira, M.~Kagan, L.~Mackey, B.~Nachman and A.~Schwartzman,
  \emph{{Jet-images \textemdash{} deep learning edition}},
  \href{https://doi.org/10.1007/JHEP07(2016)069}{\emph{JHEP} {\bfseries 07}
  (2016) 069} \href{https://arxiv.org/abs/1511.05190}{{\ttfamily
  arXiv:1511.05190}} [hep-ph].

\bibitem{Baldi:2016fql}
P.~Baldi, K.~Bauer, C.~Eng, P.~Sadowski and D.~Whiteson, \emph{{Jet
  Substructure Classification in High-Energy Physics with Deep Neural
  Networks}}, \href{https://doi.org/10.1103/PhysRevD.93.094034}{\emph{Phys.
  Rev. D} {\bfseries 93} (2016) 094034}
  \href{https://arxiv.org/abs/1603.09349}{{\ttfamily arXiv:1603.09349}}
  [hep-ex].

\bibitem{Lin:2018cin}
J.~Lin, M.~Freytsis, I.~Moult and B.~Nachman, \emph{{Boosting $H\to b\bar b$
  with Machine Learning}},
  \href{https://doi.org/10.1007/JHEP10(2018)101}{\emph{JHEP} {\bfseries 10}
  (2018) 101} \href{https://arxiv.org/abs/1807.10768}{{\ttfamily
  arXiv:1807.10768}} [hep-ph].

\bibitem{Barnard:2016qma}
J.~Barnard, E.~N. Dawe, M.~J. Dolan and N.~Rajcic, \emph{{Parton Shower
  Uncertainties in Jet Substructure Analyses with Deep Neural Networks}},
  \href{https://doi.org/10.1103/PhysRevD.95.014018}{\emph{Phys. Rev. D}
  {\bfseries 95} (2017) 014018}
  \href{https://arxiv.org/abs/1609.00607}{{\ttfamily arXiv:1609.00607}}
  [hep-ph].

\bibitem{Komiske:2016rsd}
P.~T. Komiske, E.~M. Metodiev and M.~D. Schwartz, \emph{{Deep learning in
  color: towards automated quark/gluon jet discrimination}},
  \href{https://doi.org/10.1007/JHEP01(2017)110}{\emph{JHEP} {\bfseries 01}
  (2017) 110} \href{https://arxiv.org/abs/1612.01551}{{\ttfamily
  arXiv:1612.01551}} [hep-ph].

\bibitem{Kasieczka:2017nvn}
G.~Kasieczka, T.~Plehn, M.~Russell and T.~Schell, \emph{{Deep-learning Top
  Taggers or The End of QCD?}},
  \href{https://doi.org/10.1007/JHEP05(2017)006}{\emph{JHEP} {\bfseries 05}
  (2017) 006} \href{https://arxiv.org/abs/1701.08784}{{\ttfamily
  arXiv:1701.08784}} [hep-ph].

\bibitem{Macaluso:2018tck}
S.~Macaluso and D.~Shih, \emph{{Pulling Out All the Tops with Computer Vision
  and Deep Learning}},
  \href{https://doi.org/10.1007/JHEP10(2018)121}{\emph{JHEP} {\bfseries 10}
  (2018) 121} \href{https://arxiv.org/abs/1803.00107}{{\ttfamily
  arXiv:1803.00107}} [hep-ph].

\bibitem{Choi:2018dag}
S.~Choi, S.~J. Lee and M.~Perelstein, \emph{{Infrared Safety of a Neural-Net
  Top Tagging Algorithm}},
  \href{https://doi.org/10.1007/JHEP02(2019)132}{\emph{JHEP} {\bfseries 02}
  (2019) 132} \href{https://arxiv.org/abs/1806.01263}{{\ttfamily
  arXiv:1806.01263}} [hep-ph].

\bibitem{Guest:2016iqz}
D.~Guest, J.~Collado, P.~Baldi, S.-C. Hsu, G.~Urban and D.~Whiteson, \emph{{Jet
  Flavor Classification in High-Energy Physics with Deep Neural Networks}},
  \href{https://doi.org/10.1103/PhysRevD.94.112002}{\emph{Phys. Rev. D}
  {\bfseries 94} (2016) 112002}
  \href{https://arxiv.org/abs/1607.08633}{{\ttfamily arXiv:1607.08633}}
  [hep-ex].

\bibitem{Pearkes:2017hku}
J.~Pearkes, W.~Fedorko, A.~Lister and C.~Gay, \emph{{Jet Constituents for Deep
  Neural Network Based Top Quark Tagging}},
  \href{https://arxiv.org/abs/1704.02124}{{\ttfamily arXiv:1704.02124}}
  [hep-ex].

\bibitem{Egan:2017ojy}
S.~Egan, W.~Fedorko, A.~Lister, J.~Pearkes and C.~Gay, \emph{{Long Short-Term
  Memory (LSTM) networks with jet constituents for boosted top tagging at the
  LHC}},  \href{https://arxiv.org/abs/1711.09059}{{\ttfamily arXiv:1711.09059}}
  [hep-ex].

\bibitem{Fraser:2018ieu}
K.~Fraser and M.~D. Schwartz, \emph{{Jet Charge and Machine Learning}},
  \href{https://doi.org/10.1007/JHEP10(2018)093}{\emph{JHEP} {\bfseries 10}
  (2018) 093} \href{https://arxiv.org/abs/1803.08066}{{\ttfamily
  arXiv:1803.08066}} [hep-ph].

\bibitem{Butter:2017cot}
A.~Butter, G.~Kasieczka, T.~Plehn and M.~Russell, \emph{{Deep-learned Top
  Tagging with a Lorentz Layer}},
  \href{https://doi.org/10.21468/SciPostPhys.5.3.028}{\emph{SciPost Phys.}
  {\bfseries 5} (2018) 028} \href{https://arxiv.org/abs/1707.08966}{{\ttfamily
  arXiv:1707.08966}} [hep-ph].

\bibitem{Kasieczka:2018lwf}
G.~Kasieczka, N.~Kiefer, T.~Plehn and J.~M. Thompson, \emph{{Quark-Gluon
  Tagging: Machine Learning vs Detector}},
  \href{https://doi.org/10.21468/SciPostPhys.6.6.069}{\emph{SciPost Phys.}
  {\bfseries 6} (2019) 069} \href{https://arxiv.org/abs/1812.09223}{{\ttfamily
  arXiv:1812.09223}} [hep-ph].

\bibitem{Erdmann:2018shi}
M.~Erdmann, E.~Geiser, Y.~Rath and M.~Rieger, \emph{{Lorentz Boost Networks:
  Autonomous Physics-Inspired Feature Engineering}},
  \href{https://doi.org/10.1088/1748-0221/14/06/P06006}{\emph{JINST} {\bfseries
  14} (2019) P06006} \href{https://arxiv.org/abs/1812.09722}{{\ttfamily
  arXiv:1812.09722}} [hep-ex].

\bibitem{Louppe:2017ipp}
G.~Louppe, K.~Cho, C.~Becot and K.~Cranmer, \emph{{QCD-Aware Recursive Neural
  Networks for Jet Physics}},
  \href{https://doi.org/10.1007/JHEP01(2019)057}{\emph{JHEP} {\bfseries 01}
  (2019) 057} \href{https://arxiv.org/abs/1702.00748}{{\ttfamily
  arXiv:1702.00748}} [hep-ph].

\bibitem{Cheng:2017rdo}
T.~Cheng, \emph{{Recursive Neural Networks in Quark/Gluon Tagging}},
  \href{https://doi.org/10.1007/s41781-018-0007-y}{\emph{Comput. Softw. Big
  Sci.} {\bfseries 2} (2018) 3}
  \href{https://arxiv.org/abs/1711.02633}{{\ttfamily arXiv:1711.02633}}
  [hep-ph].

\bibitem{dgcnn}
Y.~Wang, Y.~Sun, Z.~Liu, S.~E. Sarma, M.~M. Bronstein and J.~M. Solomon,
  \emph{Dynamic graph cnn for learning on point clouds},
  \href{https://doi.org/10.1145/3326362}{\emph{ACM Trans. Graph.} {\bfseries
  38} (2019) }.

\bibitem{weaver}
H.~Qu, ``\weaver.'' \url{https://github.com/hqucms/weaver}.

\bibitem{James_1980}
F.~James, \emph{Monte carlo theory and practice},
  \href{https://doi.org/10.1088/0034-4885/43/9/002}{\emph{Reports on Progress
  in Physics} {\bfseries 43} (1980) 1145}.

\end{thebibliography}\endgroup
\clearpage

\appendix

\section{Tracking speed optimisation and randomisation}
\label{app:trkopt}

\subsection{Speed optimization}
\label{sec:ftc_speed}
The method  for the computation of the covariance matrix, presented in Section~\ref{sec:ftc}, involves the inversion of the covariance matrix of the measurements, $S$. This matrix can be of order larger than 100 in many practical applications and becomes an obvious speed bottleneck. This problem is solved by generating a set of track covariance matrices in a transverse momentum-polar angle grid during the initialization stage. This grid is loaded in memory and then the track covariance matrices are calculated by means of a bi-linear interpolation over this grid. Care is taken to choose the grid nodes so that the geometry transitions are mapped accurately. In addition, since a linear combination of positive definite symmetric matrices is not always positive definite, we correct for the loss of positive definiteness in the extremely rare cases when this happens. A similar grid mapping the number of measurement points on the track is used to define if the track has a sufficient number of hits to be reconstructed. 

This method works well for tracks originating close to the primary interaction point, but clearly cannot describe correctly tracks that start inside the tracking volume such as, for instance, tracks from the decay of long lived particle  (such as $K^0_S$ or $\Lambda^0$). Since these track categories are a small fraction of the total, the full covariance calculation can be used without affecting much the overall simulation speed in these cases.

\subsection{Randomization}
\label{sec:ftc_random}
Once the ``true'' track parameters and their covariance matrix have been determined, pseudo-reconstructed tracks are generated by doing a Cholesky decomposition~\cite{James_1980} of the covariance matrix. This process transforms a symmetric positive definite matrix, $C$,  in the product of  an upper diagonal matrix, $U$, and its transposed: $C = ~ U^t\,\cdot U$. A vector, $\vec{r}$, of five Gaussian distributed random numbers with mean zero and standard deviation equal to one is generated and used to obtain the resolution smeared track parameters, $\vec{\alpha}'$:
\begin{equation}
\label{eq:smear}
\vec{\alpha}' = \vec{\alpha} + U^t\,\vec{r}
\end{equation}
It is easily shown that the covariance of the parameters $\vec{\alpha}'$ is exactly $C$:
\begin{equation}
\label{eq:proof}
<(\vec{\alpha}'- \vec{\alpha}) (\vec{\alpha}'- \vec{\alpha})^t\,>=
U^t\,<\vec{r} \, \vec{r}^t\,>U = U^t\, I U = C
\end{equation}
where the angle brackets indicate the average over many random number extractions.

\clearpage

\end{document}